\input harvmac
\input epsf.tex
\overfullrule=0mm
\newcount\figno
\figno=0
\def\fig#1#2#3{
\par\begingroup\parindent=0pt\leftskip=1cm\rightskip=1cm\parindent=0pt
\baselineskip=11pt
\global\advance\figno by 1
\midinsert
\epsfxsize=#3
\centerline{\epsfbox{#2}}
\vskip 12pt
{\bf Fig.\the\figno:} #1\par
\endinsert\endgroup\par
}
\def\figlabel#1{\xdef#1{\the\figno}}
\def\encadremath#1{\vbox{\hrule\hbox{\vrule\kern8pt\vbox{\kern8pt
\hbox{$\displaystyle #1$}\kern8pt}
\kern8pt\vrule}\hrule}}
\def\appendix#1#2{\global\meqno=1\global\subsecno=0\xdef\secsym{\hbox{#1.}}
\bigbreak\bigskip\noindent{\bf Appendix. #2}\message{(#1. #2)}
\writetoca{Appendix {#2}}\par\nobreak\medskip\nobreak}

\def\tvi{\vrule height 12pt depth 6pt width 0pt}
\def\tv{\tvi\vrule}

\Title{T96/124, SU-4240-651}
{{\vbox {
\medskip
\centerline{{\bf Geometrical Folding Transitions of the Triangular Lattice}}
\centerline{{\bf in the Face-Centred Cubic Lattice}}
}}}
\bigskip
\centerline{M. Bowick\footnote{}{\kern -20pt email: bowick@npac.syr.edu  
golinelli,guitter@spht.saclay.cea.fr 
mori@monet.phys.s.u-tokyo.ac.jp},}
\medskip
\centerline{\it Physics Department, Syracuse University,}
\centerline{\it Syracuse NY 13244-1130, USA}
\medskip
\centerline{O. Golinelli, E. Guitter,}
\medskip
\centerline{ \it CEA, Service de Physique Th\'eorique de Saclay,}
\centerline{ \it F-91191 Gif sur Yvette Cedex, France}
\medskip
\centerline{S. Mori}
\medskip
\centerline{\it Department of Physics, Graduate School of Science,}
\centerline{\it University of Tokyo, Hongo 7-3-1}
\centerline{\it Bunkyo-ku, Tokyo 113, Japan}
\baselineskip=12pt
\vskip .5in
 
We study the folding of the regular two-dimensional triangular lattice
embedded in the regular three-dimensional Face-Centred Cubic lattice,
a discrete model for the crumpling of membranes.  Possible folds are
complete planar folds, folds with the angle of a regular tetrahedron
($71^\circ$) or with that of a regular octahedron ($109^\circ$).  We
study this model in the presence of a {\it negative} bending rigidity
$K$, which favours the folding process. We use both a cluster variation
method (CVM) approximation and a transfer matrix approach.  The system
is shown to undergo two separate geometrical transitions with
increasing $\vert K\vert$: a first discontinuous transition separates
a phase where the triangular lattice is preferentially wrapped around
octahedra from a phase where it is preferentially wrapped around
tetrahedra.  A second continuous transition separates this latter
phase from a phase of complete folding of the lattice on top of a
single triangle.

\noindent
\Date{11/96}
 
\nref\JERU{``Statistical Mechanics of Membranes and Surfaces,'' D.R. Nelson, 
T. Piran and S. Weinberg eds, Proceedings of the fifth Jerusalem Winter 
School for Theoretical Physics (World Scientific, Singapore, 1989).}
\nref\LESH{``Fluctuating Geometries in Statistical Mechanics and Field
Theory,'' F. David, P. Ginsparg and J. Zinn-Justin eds; Les Houches
Session LXII (Elsevier Science, The Netherlands, 1996) 
(http://xxx.lanl.gov/lh94).}
\nref\DG{F. David and E. Guitter, Europhys. Lett. {\bf 5} (1988) 709.}
\nref\KJ{Y. Kantor and M.V. Jari\'c, Europhys. Lett. {\bf 11} (1990) 157.}
\nref\DIG{P. Di Francesco and E. Guitter, Europhys. Lett. {\bf 26} (1994)
455 (cond-mat/9402058).}
\nref\DIGG{P. Di Francesco and E. Guitter, Phys. Rev. E {\bf 50}
(1994) 4418 (cond-mat/9406041).}
\nref\BAX{R.J. Baxter, J. Math. Phys. {\bf 11} (1970) 784 and
J. Phys. A: Math. Gen. {\bf 19} (1986) 2821.}
\nref\RES{N. Yu. Reshetikhin, J. Phys. A {\bf 24} (1991) 2387.}
\nref\BLNI{H.W.J. Bl\"{o}te and B. Nienhuis, Phys. Rev. Lett. {\bf
72} (1994) 1372.}  
\nref\HURU{D.A. Huse and A.D. Rutenberg,
Phys. Rev. B {\bf 45} (1992) 7536.}  
\nref\CGP{E. Cirillo, G. Gonnella and A. Pelizzola, Phys. Rev. E {\bf
53} (1996) 1479 (hep-th/9507161).}
\nref\BDGG{M. Bowick, P. Di Francesco, O. Golinelli and E. Guitter,
Nucl. Phys. {\bf B450} [FS] 463 (1995) (cond-mat/9502063).}
\nref\CGPII{E. Cirillo, G. Gonnella and A. Pelizzola, Phys. Rev. E
{\bf 53} (1996) 3253 (hep-th/9512069).}
\nref\BDGGII{M. Bowick, P. Di Francesco, O. Golinelli and E. Guitter, 
``Discrete Folding'', to appear in the Proceedings of the 4th Chia
Meeting on ``Condensed Matter and High-Energy Physics'', September
4-8, 1995 (World Scientific, Singapore) (cond-mat/9610215).}  
\nref\ABNE{F.F. Abraham and D.R. Nelson, J. Phys. France {\bf 51}
(1990) 2653.}
\nref\K{R. Kikuchi, J. Chem. Phys. {\bf 60} (1974) 1071.}
\nref\MO{T. Morita, Prog. Theor. Phys. {\bf 103} (1984) 103.}
\nref\An{G. An, J. Stat. Phys. {\bf 52} (1988)  727.}
\nref\DGM{P. Di Francesco, E. Guitter and S. Mori, ``Folding of the
triangular lattice with quenched random bending rigidity," to appear in
Phys. Rev. E (cond-mat/9607077).}
\nref\Baxbook{R.J. Baxter, Exactly Solved Models in Statistical Mechanics
(Academic Press, 1982).}
\nref\BREZ{C. Brezinski and M. Redivo Zaglia, Extrapolation methods
(Elsevier North-Holland, 1991).}

\newsec{Introduction}

The statistical mechanics of membranes and flexible two-dimensional
interfaces is an extremely rich and rather challenging field
[\xref\JERU,\xref\LESH].  Of fundamental importance is the phase
diagram of such systems and the precise nature of the remarkable
orientationally ordered phases that they possess.  Faced with a
difficult problem it is often useful to explore extreme limits. One
such limit is that of a flexible phantom fixed-connectivity membrane
with infinite elastic stiffness. In this limit the in-plane ``phonon''
degrees of freedom are frozen, leaving only out-of-plane ``bending''
modes.  In a $1/d$ expansion (where $d$ is the dimension of the
embedding space) such a membrane exhibits a continuous crumpling
transition at a positive critical bending rigidity \DG.

Discretising this type of membrane as a triangular lattice leads to a
model of the statistical mechanics of folding.  Finally one can
discretise the space of allowed local folds to obtain a completely
discrete ``vertex'' model.  The simplest case is planar folding, in
which the local folds are no-folds or complete folds
\refs{\KJ{--}\DIGG}.  This corresponds to folding the triangular
lattice onto itself. Planar folding admits many equivalent
representations, each of them interesting.  It may be formulated as a
11-vertex model equivalent to the three-colouring problem of a
triangular lattice \refs{\DIG,\DIGG}. For zero bending rigidity this
model was solved by Baxter, in its dual formulation, as the
combinatorics of edge-colourings of the honeycomb lattice \BAX.  It
may also be mapped to the fully packed loop model with fugacity 2
describing non-intersecting loops on a honeycomb lattice such that
every vertex is covered by a loop \refs{\RES,\BLNI}. Finally an
interpretation of the $Z_3$ colour variable as a Potts spin living on
the centres of the bonds maps planar folding to a description of the
ground state of the three-state antiferromagnetic Potts model on the
Kagom\'e lattice \HURU.  Adding a positive bending rigidity to the
planar folding model one finds a first-order phase transition to a
completely unfolded phase \refs{\DIGG,\CGP}.

Since planar folding is very restrictive, it is natural to consider
enlarging the discrete space of allowed folds. In reference \BDGG\
planar folding was generalised to a particular three-dimensional
folding allowing four types of discrete folds.  This corresponds to
choosing the embedding space to be a three-dimensional Face-Centred
Cubic lattice. This 3-d folding model has a formulation as a 96-vertex
statistical mechanics model and, once again, undergoes a first-order
transition to an ordered phase at a definite positive bending rigidity
\refs{\CGPII,\BDGGII}. It also admits a formulation as a dressed
three-colouring problem \BDGG.

In this paper we extend the study of the 96-vertex model to {\it
negative} bending rigidity. In such a model folding is energetically
favoured. Negative bending rigidity might in principle be realisable
in a polymerised membrane with strong next-to-nearest neighbour
interactions \ABNE\ or for a membrane in a sufficiently bad solvent.

Utilising both the cluster variation method and a transfer matrix
formalism we find two distinct geometrical phase transitions at
negative bending rigidity. As $K$ becomes more negative starting from
zero, there is a first-order transition from a phase of preferential
octahedral wrapping to a phase with preferential tetrahedral wrapping,
followed by a continuous piling transition to a single plaquette.  We
determine a set of three order parameters that completely characterise
the separate phases observed.
 
The outline of the paper is as follows.  In Section 2 we review the
construction of the 96-vertex model. In Section 3 we discuss four
natural submodels of the 96-vertex model {---} these correspond to
different restrictions on the allowed folds. We also introduce a
duality transformation which maps any allowed folding configuration to
another allowed folding configuration.  In Section 4 we study the
statistics of folding for negative bending rigidity and introduce the
three geometrical order parameters.  The phase diagram is determined
by the Cluster Variation Method (CVM) and by a transfer matrix method.
In Section 5 we discuss the incorporation of external fields.  Finally
we summarise our results and their significance in Section 6.

\newsec{The 96-vertex model}

\subsec{General definition of the folding model}

In this section, we first recall the model of discrete folding of the
regular 2-dimensional triangular lattice embedded in the regular
3-dimensional Face-Centred Cubic (FCC) lattice. A detailed
derivation of the model can be found in \BDGG . As explained below,
the model is easily expressible in terms of {\it two} $Z_2$ {\it
variables} $\sigma$ and $z$, defined on the faces of the triangular
lattice. The associated domain walls indicate the different types of
folds.  The spin variables $\sigma$ and $z$ are subject to {\it two basic
folding rules}, leading to a re-expression of the model as a {\it
96-vertex} model on the triangular lattice.

\fig{The Face-Centred Cubic lattice viewed as a packing of 3-d space
with octahedra and tetrahedra.}{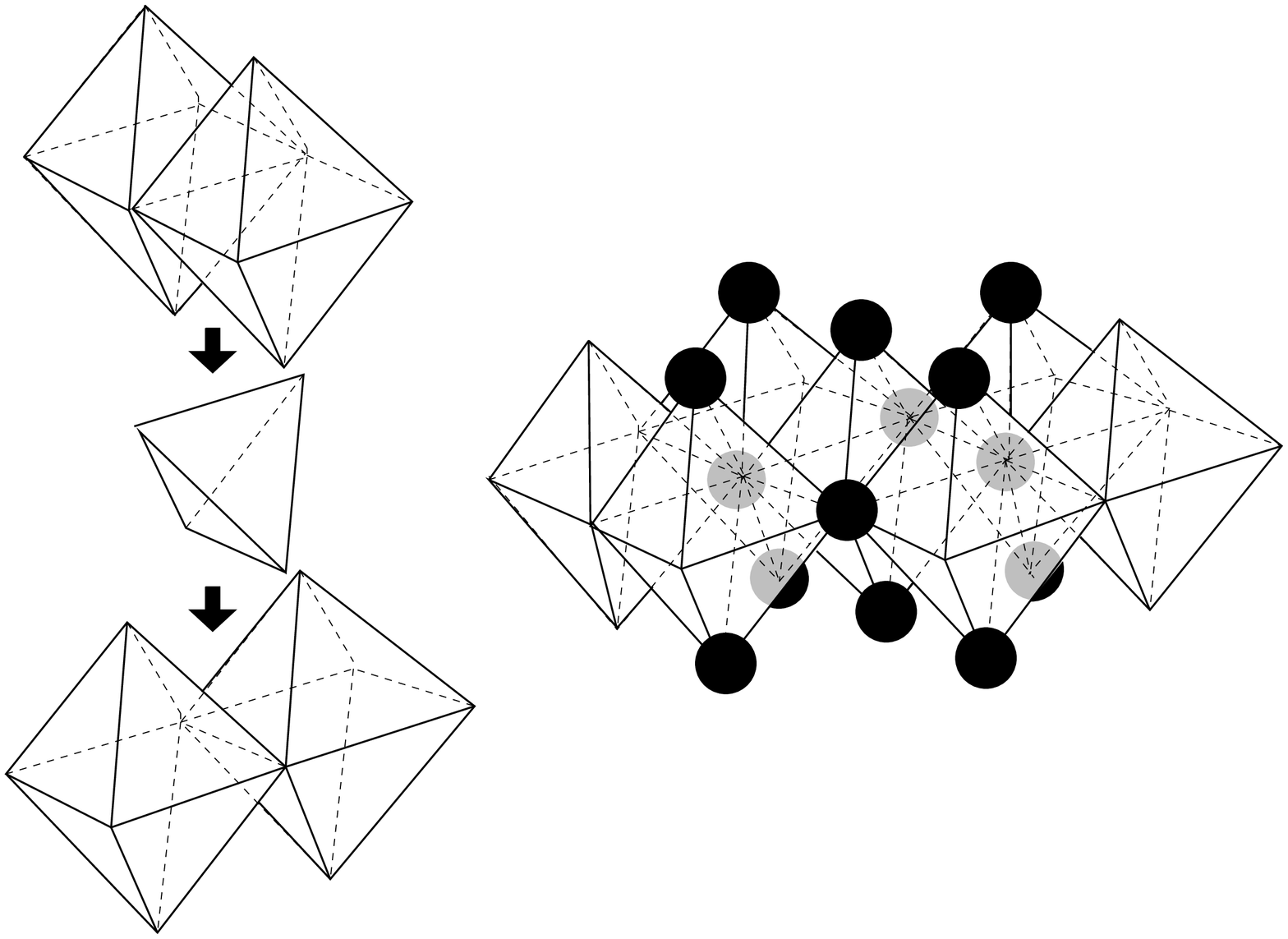}{4.0truein} 
\figlabel\fcc
\fig{An example of folding in the FCC lattice of an elementary hexagon
cut out from the triangular lattice. Nearest neighbours in the hexagon
remain nearest neighbours after embedding in the FCC lattice.}
{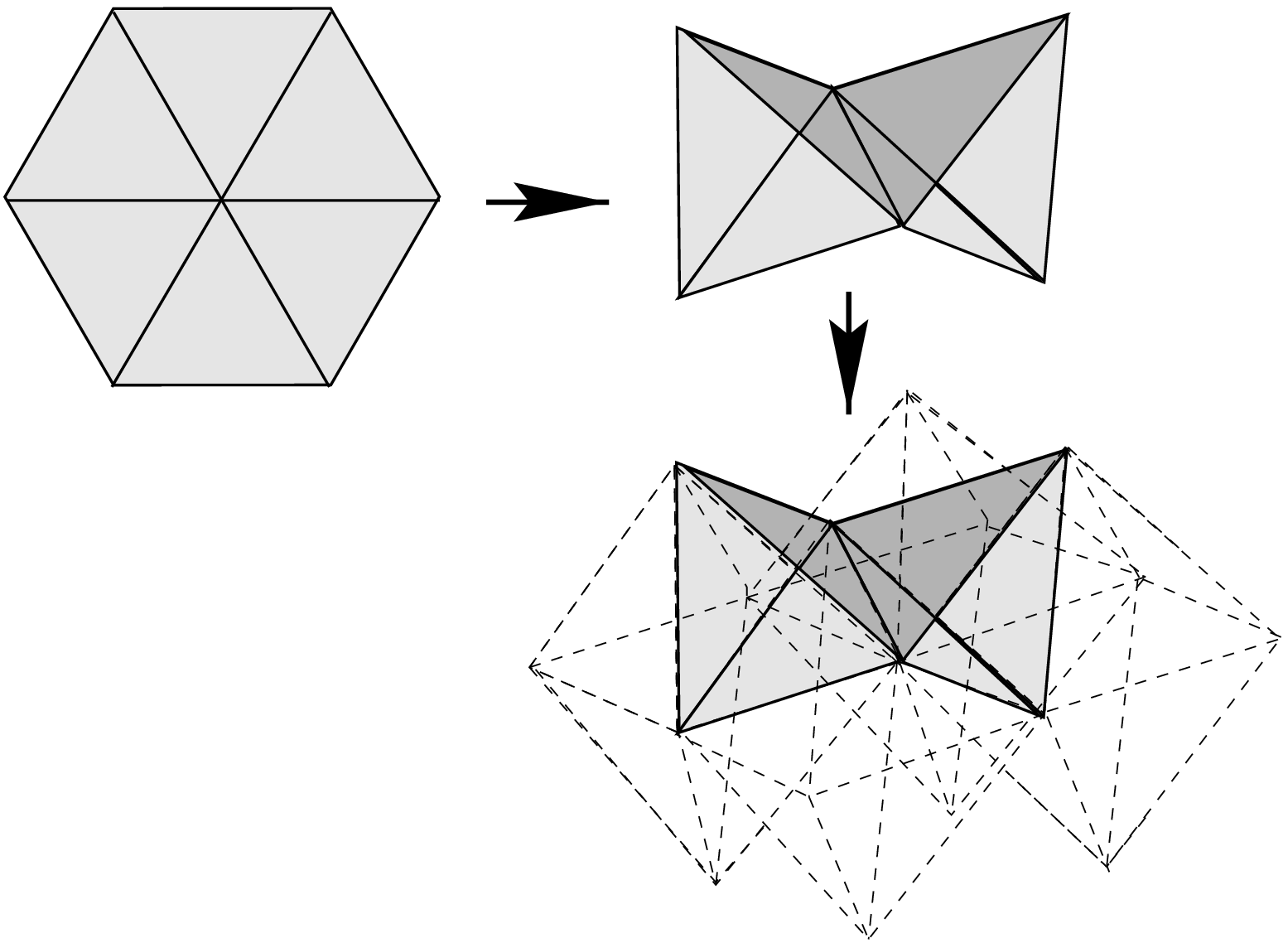}{8.truecm} 
\figlabel\foldex 
A folding of the triangular lattice in the FCC lattice is simply a
mapping which assigns to each vertex of the triangular lattice a
vertex in the FCC lattice, with the constraint that neighbouring
vertices on the triangular lattice are sent to nearest neighbours in
the FCC lattice.  The FCC lattice can be viewed as consisting of a
filling of space by regular octahedra complemented by regular
tetrahedra (see Fig. \fcc ).  Nearest-neighbour triplets in the FCC
lattice thus form equilateral triangles.  In the folding process each
elementary triangle of the triangular lattice is mapped onto one of
these equilateral triangles of the FCC lattice.  Fig. \foldex\ shows 
an example of folding of a piece of triangular lattice
consisting of an elementary hexagon excised from the lattice.

\fig{The four possible folding angles between two adjacent triangles.
The neighbour of the dark triangle may (i) occupy the symmetric
position in the same plane (no fold), (ii) be itself on top of the
dark triangle (complete fold), (iii) lie on the same tetrahedron
(i.e. form an acute angle), (iv) lie on the same octahedron (i.e. form
an obtuse angle).  }{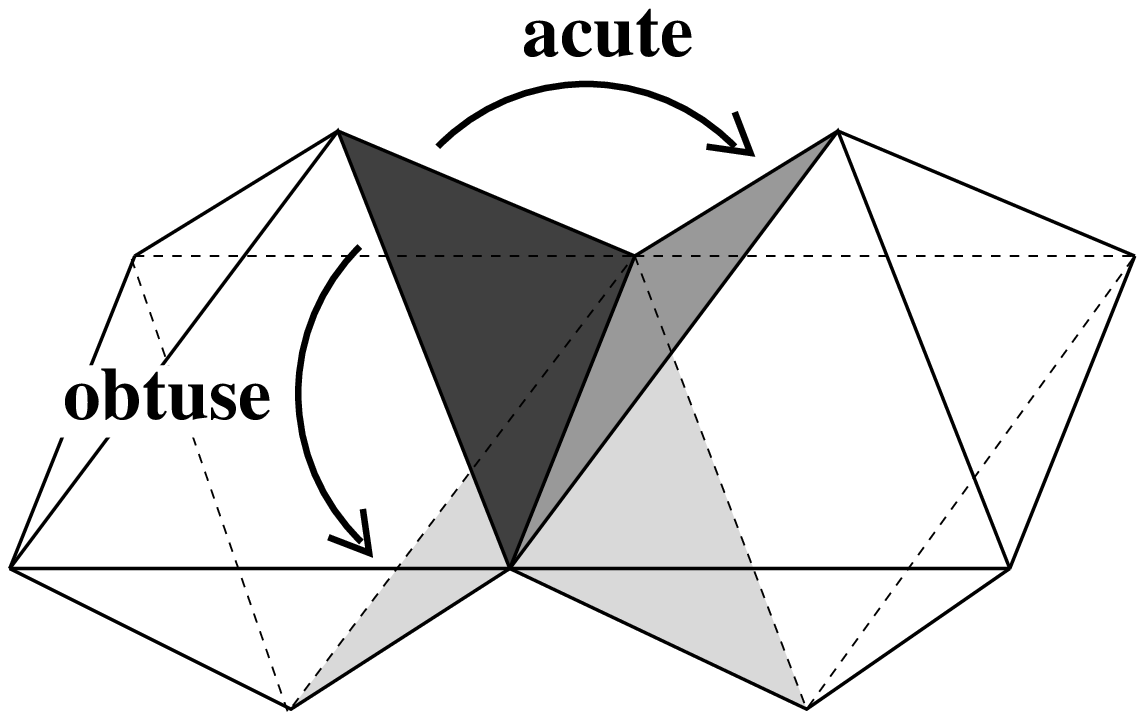}{6.truecm} \figlabel\acutob 
After folding, two adjacent triangles form some relative
angle, i.e.  their common link serves as a hinge and may
be (partially) folded or not.  There are four possible relative
positions for two given neighbouring triangles, as depicted in
Fig. \acutob , measured simply by the angle $\theta$ of the
corresponding fold:
\item{(i)} $\theta=180^\circ$ {---} no fold: the triangles are side by side;
\item{(ii)} $\theta=0^\circ$ {---} complete fold: the triangles are on top
of each other;
\item{(iii)} $\theta=\arccos(1/3)\sim 71^\circ$ {---} fold with acute
angle: the two triangles lie on the same tetrahedron, and
\item{(iv)} $\theta=\arccos(-1/3)\sim 109^\circ$ {---} fold with obtuse
angle: the triangles lie on the same octahedron.  
\par 
It has been shown in \BDGG\ that these four possible types of folds
are actually organised as the domain walls for two $Z_2$ variables
$\sigma=\pm 1$ and $z=\pm 1$ defined on each face of the triangular
lattice. The relative value of both $z_2/z_1$ and $\sigma_2/\sigma_1$
for two neighbouring triangles indicate which type of fold they form,
with the correspondence displayed in Table~I below.  The domain walls
for the $z$ variable are the location of the folds which are either
acute or obtuse, whereas those for the $\sigma$ variable are the
location of the folds which are either complete or obtuse.  The
superposition of these two types of domain wall fixes the folding
state of all the links, specifying the folding state of the lattice
(up to a global orientation).
%
\midinsert
$$\vbox{\offinterlineskip
\halign{\tv\quad # & \quad\tv \quad
# & \quad \tv \quad  # & \quad \tv #\cr
\noalign{\hrule}
\tvi $z_2/z_1$ & $\sigma_2/\sigma_1$ & angle &\cr
\noalign{\hrule}
\tvi  $\phantom{-}1$ &  $\phantom{-}1$  &    no~fold  &\cr
\tvi  $\phantom{-}1$ &  $-1$  &    complete~fold  &\cr
\tvi  $-1$ &  $\phantom{-}1$  &    acute~fold  &\cr
\tvi  $-1$ &  $-1$  &    obtuse~fold  &\cr
\noalign{\hrule} }} $$
\par\begingroup\parindent=0pt\leftskip=1cm\rightskip=1cm\parindent=0pt
\baselineskip=11pt {\bf Table I:} The relative folding state of two
neighbouring triangles according to their relative values of $z_i=\pm
1$ and $\sigma_i=\pm 1$ ($i=1,2$).  
\par 
\endgroup\par 
\endinsert

Note, of course, that the two global symmetries $\sigma \to -\sigma$
and $z \to -z$, independently, leave the folding invariant.

\subsec{The two basic folding rules}

\fig{The six $z_i$ and $\sigma_i$ variables around a given vertex, and
the colours $c_i$ of the interior links.}{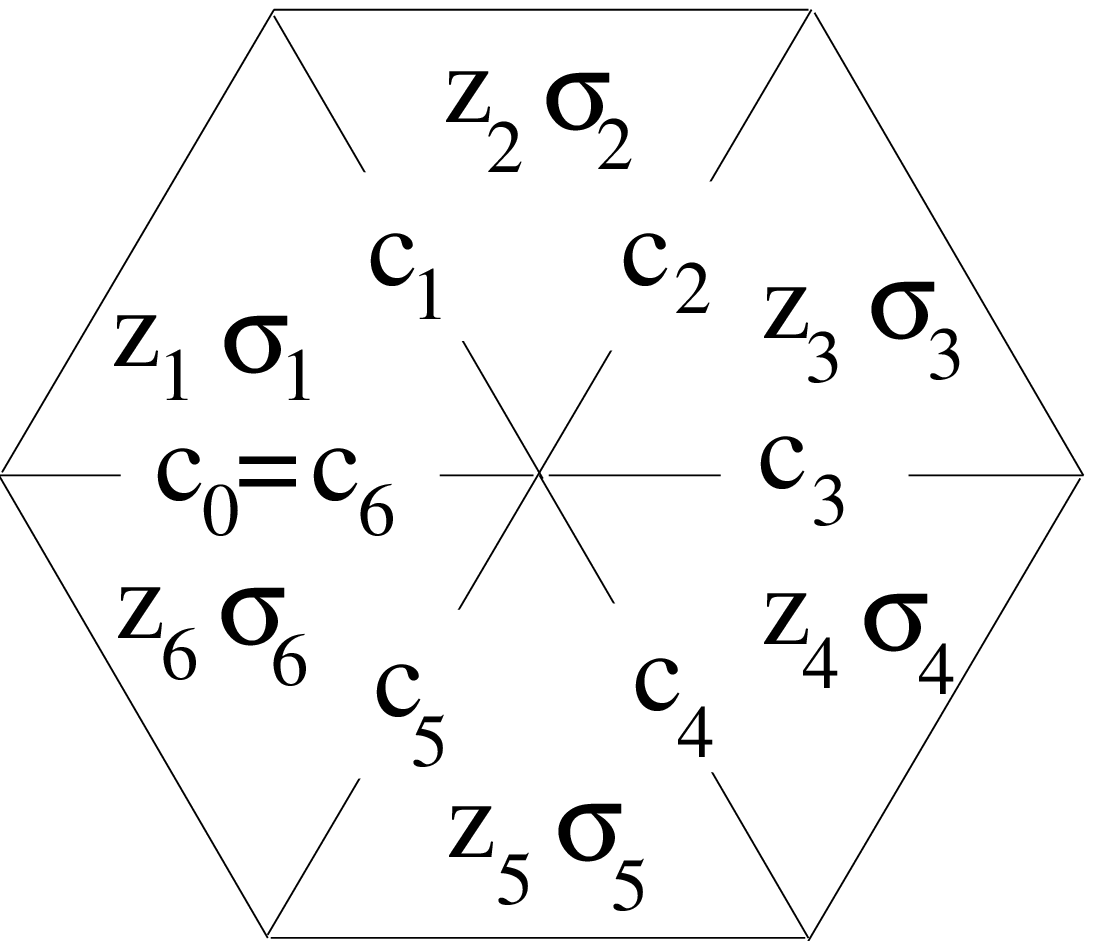}{4.truecm}
\figlabel\spifig With the above definitions, the $\sigma$ and $z$
variables on neighbouring triangles are not totally independent. This
can be seen by considering the folding of any elementary hexagon made
of six neighbouring triangles in the lattice. The relative positions of
the six surrounding vertices of the hexagon can be traced back from
the folds of its six internal links. After a complete turn around the
hexagon, the same absolute position must be found for any of these six
vertices. This leads to two basic folding rules for the values
$\sigma_i$ and $z_i$ ($i=1,\cdots,6$) on these six triangles (see
Fig. \spifig ).
 
The first rule involves the variable $\sigma$ only and states
that:

\eqn\firstrule{\sum_{i=1}^6 \sigma_i = 0 \ {\rm mod}\ 3 .}

This rule is easily understood in terms of 3--colourability of the
links of the triangular lattice as follows.  When embedded in the FCC
lattice, the links of the triangular lattice must belong to one of the
three crystalline planes (200), (020) or (002) which cross the
octahedra by their three square bases. If we assign a colour to each of
these three planar directions and colour the links of the FCC lattice
correspondingly, this induces a 3--colouring of the triangular lattice
links with three colours ($c=0,1,2$ mod 3), which, moreover, have to be
distinct on the three edges of the elementary triangles.  One can
easily check that the colours $c_i$ of the six interior links of a
hexagon (see Fig. \spifig ) are directly linked to the $\sigma_i$
variables by: \eqn\spico{c_j-c_i=\sum_{k=i+1}^j \sigma_k \ {\rm mod}\
3 .}  For instance, if $\sigma_2/\sigma_1=-1$, then
$c_2-c_0=\sigma_1+\sigma_2=0$ since a complete or an obtuse fold puts
the link 2 in the same crystalline plane as the link 0.  The first
basic folding rule simply states that the same colour must be found
after a complete turn around the hexagon, i.e. that $c_0=c_6$.

This first rule on the $\sigma$ variable is not sufficient to ensure
that the $(z_i,\sigma_i)$ do correspond to an actual folded state of
the hexagon. This is because the colour only partially specifies the
actual orientation of the links in 3-dimensional space. A second
basic folding rule involves both the $z$ and $\sigma$ variables and
can be stated as follows: \eqn\secondrule{\eqalign{&\prod_{i\in
I(c)}z_iz_{i+1}=1 \quad{\rm for}\ c=0,1,2 \cr &I(c)=\{i\ :\
\sum_{k=1}^i \sigma_k = c \ {\rm mod}\ 3 \}.\cr}} This rule states
that the number of changes of the $z$ variable (which indicate folds
which are either obtuse or acute) occurring on all interior links of a
given colour in the hexagon must be even.  The interpretation of this
second rule will be clarified in Section 3 by use of a duality
transformation. A direct interpretation based on the tetrahedral group
can be found in \BDGG .

It can be shown that the two rules \firstrule\ and \secondrule ,
applied independently to all the hexagons of the original triangular
lattice, are necessary and sufficient to define a folded state in the
FCC lattice. Of course, our definition of folding does not impose any
restrictions on possible self-intersections of the triangular lattice,
which is therefore considered as a phantom object. In the above
formulation, the folding problem can thus be understood as a {\it
constrained} system of two $Z_2$ variables $z$ and $\sigma$.  \fig{The
96 possible hexagonal configurations around a vertex of the triangular
lattice. Complete folds are represented by thick solid lines, obtuse
folds by thin solid lines, and acute folds by dashed lines.  We have
represented only the vertices which are not equivalent to each other
by a simple rotation. We indicate for each vertex the corresponding
degeneracy under rotation.  }{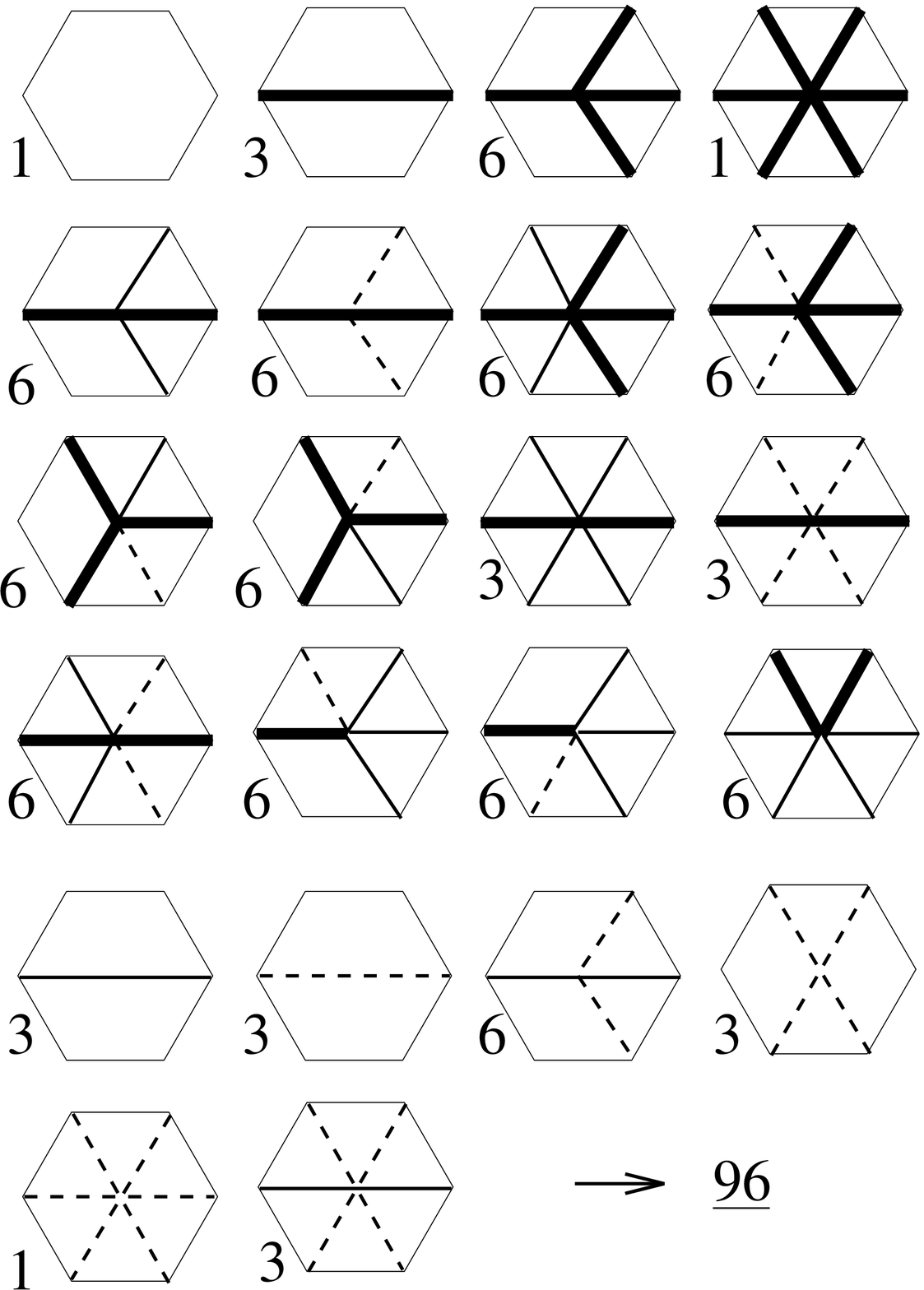}{10.truecm}
\figlabel\vertex It is interesting to work out all possible hexagonal
configurations around a vertex allowed by the two rules \firstrule\
and \secondrule .  If one forgets the global fourfold symmetry under
global reversal of all $\sigma$ or $z$ and represents only the
corresponding domain walls, i.e.  the four types of folds (i)-(iv)
discussed above, one is led to 96 possible hexagonal configurations
for any of the vertices of the triangular lattice. These are
represented in Fig. \vertex .  Our folding problem can thus be thought
of as a {\it 96-vertex model} defined on the triangular lattice.

\subsec{The Folding Energy}

For each type of fold, we introduce the folding energy:
\eqn\energ{E=K\ \cos (\theta ) ,} where $\theta$ is the angle of the
fold ($180^\circ$ for no fold, $0^\circ$ for a complete fold,
$71^\circ$ for an acute fold and $109^\circ$ for an obtuse fold). The
parameter $K$ is the bending rigidity. In the following we shall be
interested mainly in the case of {\it negative} $K$, which corresponds
to favouring folds.  In terms of the variables $\sigma_1,\sigma_2$ and
$z_1,z_2$ of the two triangles forming the fold, the energy simply
reads: \eqn\ener{E=-{K\over 3} \sigma_1\sigma_2 \,(1+2z_1z_2) .}  The
total folding energy is the sum of all elementary folding energies for
all links of the triangular lattice.

\newsec{Submodels and duality}

In this section, we first introduce four submodels of the 96-vertex
model describing four particular restricted folding problems.  These
models can be seen as particular limiting cases of the full 3-d folding
problem, where some particular {\it complete} ordering is realised in
the folding process, which correspond to a restriction of the target
space to subsets of the FCC lattice. We shall see in the next section
that similar {\it partial} orderings occur spontaneously in the
unrestricted full 3-d folding problem.  We also derive here a duality
relation for the 96-vertex model.

\subsec{Planar folding: an 11-vertex model}

A first submodel can be obtained by restricting the target space to a
single plane made of triangles in the FCC, say the crystalline plane
(111) which cuts the FCC lattice along a regular 2-dimensional
triangular lattice.  In this case, we are dealing with the pure
2-dimensional folding problem studied in
[\xref\KJ-\xref\DIGG]. Here, only complete folds or no-folds are
allowed.  The model can thus be characterised by: \eqn\planar{z=+1}
globally on all triangles of the lattice (of course the choice $+$ is
arbitrary and a solution with $z=-1$ globally would be equally
acceptable). With this constraint, the second folding rule
\secondrule\ is automatically satisfied and the problem translates
into a model with a single variable $\sigma$ constrained by the
folding rule \firstrule . This model has been studied extensively in
\DIG .  In particular, the number of folded states of a lattice of $N$
triangles is found to grow like $q^N$ with $q= \sqrt{3}
\Gamma(1/3)^{3/2}/2\pi \sim 1.208$.  In terms of vertices, the model
leads to 11 allowed vertices, which are those of the first row in
Fig. \vertex .

\subsec{Folding on a single octahedron: a 16-vertex model}

One can restrict the target space to a single octahedron in the FCC
lattice by simply allowing complete and obtuse folds only. Such folds
are characterised by a change in the value of $\sigma$ on neighbouring
triangles.  It is thus convenient to introduce a {\it staggered}
variable by dividing the original triangular lattice in the two
subsets $A$ and $B$ made of all triangles pointing up and all
triangles pointing down respectively in the original (unfolded)
triangular lattice. We define \eqn\stag{\eqalign{& \sigma_{\rm st}=
\sigma \ \ \ {\rm on\ subset\ A\, ,}\cr & \sigma_{\rm st}= - \sigma \
{\rm on\ subset\ B\, .}\cr }} The submodel can then be simply
characterised by: \eqn\octah{\sigma_{\rm st}=+1} globally on the
triangular lattice.  With this constraint, the first folding rule is
automatically satisfied since $\sum_{i=1}^6\sigma_i=0$ on all
hexagons. The second folding rule takes a particularly simple form
since Eq.\octah\ imposes that the interior links of any particular
hexagon take only two colours, say $0$ and $1$, which alternate such
that $I(0)=\{1,3,5\}$ and $I(1)=\{2,4,6\}$. The rule \secondrule\ then
translates into a single equation: \eqn\secondsimp{\prod_{i=1}^6z_i=1
.}  \fig{The 16 vertices made of complete and obtuse folds only,
characterising the folding on a single
octahedron.}{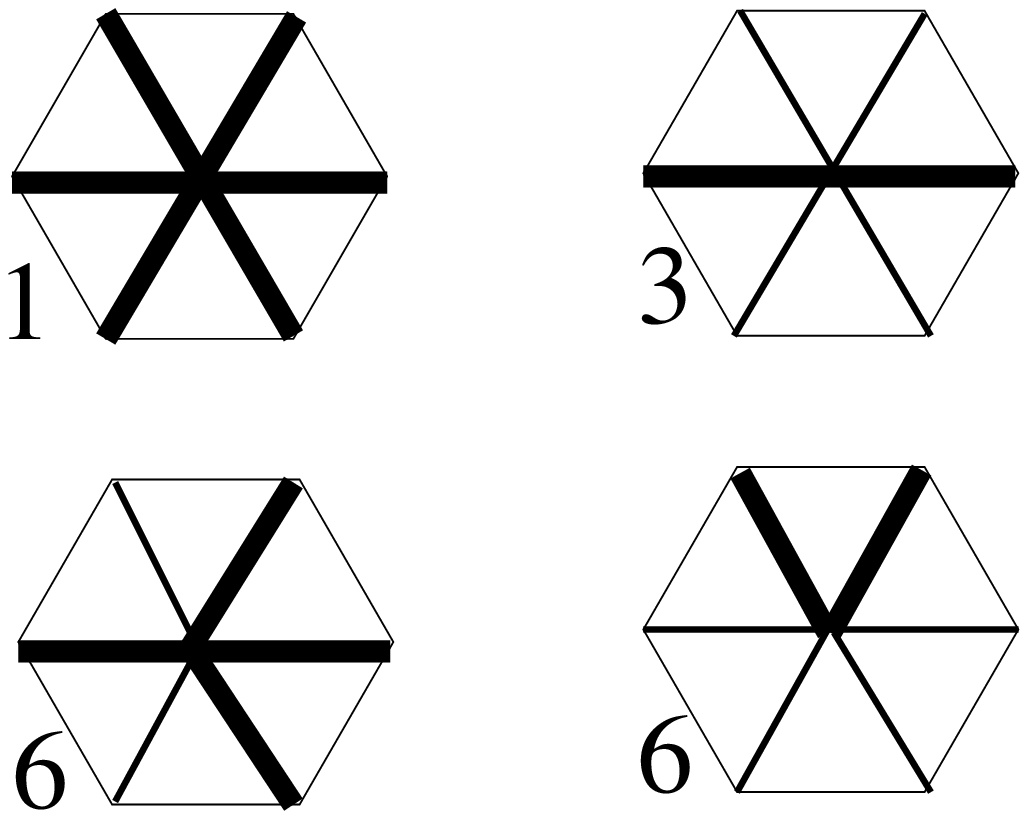}{6.truecm} \figlabel\octafig The two rules
\octah\ and \secondsimp\ leave us with the 16 vertices of
Fig. \octafig , where the thin lines (obtuse folds) indicate the
domain walls of the $z$ variable.  One can easily show that the number
of allowed folding configurations for $N$ triangles grows like $q^N$,
with $q=\sqrt{2}$. Indeed, all the entropy comes from
the $z$ degree of freedom and the constraint \secondsimp\ is solved by
introducing $Z_2$ variables $\eta=\pm 1$ on the {\it vertices} of the
triangular lattice and requiring that the variable $z$ on a given
triangle is the product of the three $\eta$ variables on the three
surrounding vertices. In this formulation the variables $\eta$ are
not constrained, leading to $2^{N/2}$ configurations for the $N/2$
vertices of the lattice.

\subsec{Folding on a single tetrahedron: an 11-vertex model}

A third submodel corresponds to the folding on a single tetrahedron,
meaning only complete and acute folds are allowed. The model can be
characterised simply by requiring that \eqn\tetrah{z=+\sigma_{\rm st
}} for all triangles in the lattice, with $\sigma$ variables
satisfying the first folding rule \firstrule .  Note that here again,
\tetrah\ automatically implies the second folding rule \secondrule\ so
that no extra constraint has to be imposed on the $\sigma$
variables. In other words, for $\sigma$ variables satisfying the first
folding rule \firstrule , the condition \eqn\seco{\prod_{i\in I(c)}
(\sigma_{\rm st})_i\ (\sigma_{\rm st})_{i+1}= \prod_{i\in I(c)}
(-\sigma_i\sigma_{i+1})=1} is automatically satisfied. This property
can be checked directly by inspection of the 22 sets of $\sigma_i$
satisfying the first folding rule. In our context, it can be
understood simply as follows: given an arbitrary hexagonal
configuration of $\sigma$ variables around a vertex, satisfying the
first basic folding rule, this configuration defines a particular {\it
planar} folded state in, say, the (111) plane.\foot{This planar folded
state and the tetrahedral folded state will be dual to each other
under the duality transformation described in the next section.}
Given an arbitrary colour $c$, the image in the FCC lattice of the central
vertex of the hexagon belongs to exactly one plane of the chosen
colour.  The (111) plane intersects this plane through a line of the
chosen colour. Successive triangles around the hexagon can lie on the
same side of this line or not. The passage from one side to the other
takes place on a link of the chosen colour $c$ and requires no fold,
characterised by no change in $\sigma$, i.e.  by a change in
$\sigma_{\rm st}$.  The number of changes in $\sigma_{\rm st}$
occurring on links of the colour $c$ thus indicates the total number of
crossings of the line when making a turn around the hexagon. This
number must clearly be even since one must get back to the same side
after one turn.  This leads to \seco .  \fig{The 11 vertices made of
complete and acute folds only, characterising the folding on a single
tetrahedron.}{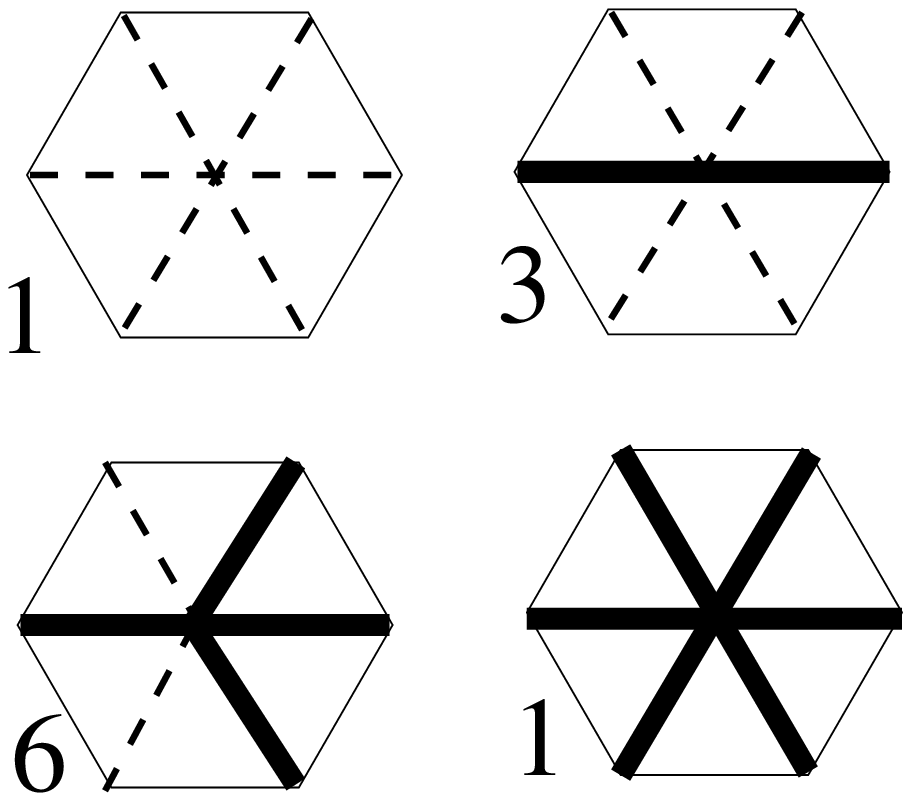}{6.truecm} \figlabel\tetrafig The submodel
of folding on a single tetrahedron is restricted to the 11 vertices
displayed in Fig.~\tetrafig , made of acute and complete folds
only. These vertices are in direct correspondence with those of the
planar 11-vertex model by replacing no-folds by acute folds. This is a
particular example of the duality transformation discussed below.

\subsec{Folding on a single triangle: a 1-vertex model}

Finally, the three submodels above share a common vertex with all
folds being complete, characterised by $z=\sigma_{\rm st}=+1$
everywhere.  This trivial case describes a state of complete folding
of the triangular lattice on top of a single triangle.

The above results are summarised in Table II, which lists for each
submodel the corresponding rules for the $\sigma$ and $z$ variables,
the restricted target space and the associated number of vertices.
%
\midinsert
$$\vbox{\offinterlineskip
\halign{\tv \ # & \tv \  
# &  \tv \  # &  \tv \  # &  \tv \ #  \tv \cr
\noalign{\hrule}
\tvi \hfill $\sigma$ variable \hfill & \hfill $z$ variable \hfill &
\hfill target space \hfill & \hfill vertices \hfill &\hfill dual model
\hfill \cr
\noalign{\hrule}
\tvi  $\displaystyle \sum_{i=1}^6\sigma_i=0$ mod 3 & \hfill $z=+1$
\hfill & \hfill single plane \hfill  & \hfill 11 \hfill & \hfill  
single tetrahedron \hfill  
\cr
\tvi  \hfill $\sigma_{\rm st}=+1$\hfill &  $\displaystyle \prod_{i=1}^6z_i=1$  
&\hfill   single octahedron \hfill  & \hfill 16 \hfill & \hfill
self-dual \hfill
\cr
\tvi  $\displaystyle \sum_{i=1}^6\sigma_i=0$ mod 3 & \hfill 
$z=+\sigma_{\rm st}$
\hfill & \hfill single tetrahedron \hfill  & \hfill 11 \hfill & \hfill
single plane \hfill  \cr
\tvi \hfill $\sigma_{\rm st}=+1$ \hfill & \hfill $z=+1$ \hfill &  
\hfill single triangle\hfill   & \hfill 1 \hfill & \hfill self-dual \hfill \cr
\noalign{\hrule} }} $$
\par\begingroup\parindent=0pt\leftskip=1cm\rightskip=1cm\parindent=0pt
\baselineskip=11pt
{\bf Table II:} 
Different submodels of the 96-vertex model.
\par
\endgroup\par
\endinsert

\subsec{The duality transformation}

Given any folded state characterised by its $\sigma$ and $z$ variables
satisfying the two basic folding rules around all hexagons, we have
seen that the first rule automatically implies Eq.~\seco . Multiplying
this equation with the second folding rule on the $z$ variables, we
get: \eqn\dualrel{\prod_{i\in I(c)}(z\sigma_{\rm st})_i (z\sigma_{\rm
st})_{i+1}=1\, ,} indicating that the global transformation
\eqn\dualtr{(\sigma,z)\to(\sigma,w\equiv z\sigma_{\rm st})} maps any
folded state of the triangular lattice to another equally acceptable
folded state.  We shall refer to the above transformation as the
duality transformation of the 96-vertex model. It is clearly an
involution, since $\sigma_{\rm st}^2=1$, and corresponds geometrically
to a global interchange of all acute folds and no-folds.

It is interesting to look at the behavior of the different submodels
of the previous subsection under the duality transformation.  The
16-vertex model is self-dual in the strong sense that any of its
vertices is invariant under the duality transformation. The 11-vertex
model describing the planar folding problem and that describing
tetrahedral folding are dual to each other. In particular the study of
reference \DIG\ on the 2-d folding problem applies to the tetrahedral
folding problem as well.  The duality transformations of the different
submodels are shown in Table II.

Finally, let us discuss again the meaning of the second folding rule
\secondrule .  By duality, this second rule is totally equivalent to
\dualrel .  Now \dualrel\ simply expresses that the number of changes
in the variable $z\sigma_{\rm st}$ occurring on the links of a given
colour must be even. Such changes account for obtuse folds or
no-folds, which are precisely the two ways of {\it crossing} the
crystalline plane of the given colour.  The number of such crossings
on a loop must clearly be even, implying Eq. \dualrel .

To conclude this section, we would like to mention that the duality
transformation is a particular case of a more general group
transformation for the allowed $z$ configurations {\it at a fixed}
$\sigma$ {\it global configuration} satisfying the first folding rule
\firstrule .  Indeed, given two folded states $(\sigma,z)$ and
$(\sigma,z^\prime)$ defined on the entire lattice with the same
$\sigma$, the configuration $(\sigma,zz^\prime)$ also defines an
allowed folded state, as is clear from the multiplicative nature of
the second folding rule \secondrule .  The duality transformation
simply corresponds to the allowed configuration $z^\prime=\sigma_{\rm
st}$.

\newsec{Geometrical folding transitions}
 
In this section, we will study the statistics of folding in the case
of {\it negative} bending rigidity $K$.  We will be interested
primarily in the three following average values: \eqn\OP{\eqalign{
&O\equiv \langle \sigma_{\rm st} \rangle \, , \cr &T\equiv \langle
z\sigma_{\rm st} \rangle \, ,\cr &P\equiv \langle z \rangle \, ,\cr}}
with of course $-1\le P,O,T\le 1$.  The average is taken over both all
folded configurations and all triangles in the lattice. Following
Table II, we will refer to $P$, $O$ and $T$ as the planar, octahedral
and tetrahedral order parameter respectively, since $P=1$
(respectively $O=1$ and $T=1$) indicate a situation of pure planar
folding (respectively folding on a single octahedron and folding on a
single tetrahedron).  A non-zero value of $P$ will thus be the
signature that the triangular lattice preferentially stays within
planes of the FCC lattice.  Similarly, we will interpret a non-zero
value of $O$ (respectively $T$) as a tendency for the triangular
lattice to be preferentially wrapped around octahedra (respectively
tetrahedra).

We will also consider two-point functions such as: \eqn\TWO{\eqalign{
&O2\equiv \langle (\sigma_{\rm st})_1(\sigma_{\rm st})_2 \rangle \,
,\cr &T2\equiv \langle (z\sigma_{\rm st})_1(z\sigma_{\rm st})_2
\rangle \, ,\cr &P2\equiv \langle z_1z_2 \rangle \, ,\cr}} for the
correlations on two neighbouring triangles in the lattice.

\subsec{The CVM approximation}

The Cluster Variation Method (CVM) is an approximation of the exact
variational formulation of the problem [\xref\K-\xref\An].  Consider
our system of variables $z$, $\sigma$ on the lattice and the energy
\ener\ summed over all pairs of neighbouring triangles.  In terms of a
density matrix $\rho(\{\sigma,z\})$ for each {\it allowed} global
folding configuration $\{\sigma,z\}$, the exact free energy \ $F$ of
the system can be obtained from:
\eqn\free{F=\Big[\sum_{\{\sigma,z\}}\rho
(\{\sigma,z\})[E(\{\sigma,z\}) +\ln \rho(\{\sigma,z\})]\Big]_{\rm min}
.}  Here we absorb the temperature factors $k_BT$ in the definition of
$E$ and $F$.  The subscript {\it min} means that the expression is
taken at its minimum with respect to $\rho(\{\sigma,z\})$, with the
normalisation constraint: \eqn\norsig{\sum_{\{\sigma,z\}}
\rho(\{\sigma,z\})=1.}  This is the well-known variational principle.
Average values are calculated according to: \eqn\expect{\langle
A(\{\sigma,z\}) \rangle = \sum_{\{\sigma,z\}}\rho_{\rm
min}(\{\sigma,z\})A(\{\sigma,z\}),} where $\rho_{\rm
min}(\{\sigma,z\})$ is the density at the minimum of the free energy
functional \free .

The CVM approximation is obtained by truncating the generalised
cumulant expansion for the entropy part of \free , $S \equiv -
\sum_{\{\sigma,z\}} \rho(\{\sigma,z\})$ $\ln \rho(\{\sigma,z\})$, to a
set of clusters with a fixed maximal size. Here we choose this set to
be made of all elementary hexagons in the system and all their
subclusters, i.e. we take the elementary hexagons, made of six
neighbouring triangles, as the largest clusters.  We thus use the
truncated entropy: \eqn\truncent{ S_{\rm CVM}= \sum_{\rm hexagons\ H}
S_H-\sum_{\rm pairs\ P} S_P + \sum_{\rm triangles\ T} S_T.}  Here we
call $S_H=-\rho_H\ln(\rho_H)$ the entropy obtained from the reduced
density $\rho_H$ for a given hexagon, and similarly for $S_P$ and
$S_T$.  Due to the overlap of neighbouring hexagons, we must subtract
the contribution of pairs of neighbouring triangles to avoid
overcounting. Similarly, we must add back the contribution of single
triangles to account for the overlap of neighbouring pairs.  We
assume, moreover, that the reduced density $\rho_H$ is the same for
each hexagon, i.e. that the distribution is translationally invariant.
The variational principle will then be applied to this common reduced
density matrix associated with the hexagons, which we write as:

\eqn\hex{\rho_{6}(\sigma_{1},\sigma_{2},\sigma_{3},\sigma_{4},\sigma_{5},
\sigma_{6},z_{1},z_{2},z_{3},z_{4},z_{5},z_{6})} 

\noindent with the notation of Fig.~\spifig\ for the $\sigma$ and $z$
variable on a hexagon. In particular, we choose a numbering such
that odd indices refer to triangles of subset A and even indices to
triangles of subset B.  The above reduced density matrix represents
the probability for one hexagon to have fixed values of $\sigma$ and
$z$. It is non-zero only for the allowed sets of $\sigma$ and $z$
satisfying the two basic folding rules, and it is normalised to one.
We also introduce the two-point and one-point functions to account for
$S_P$ and $S_T$ in \truncent : 
\eqn\trace{\eqalign{
\rho_{2}(\sigma_{1},\sigma_{2},z_{1},z_{2})\equiv {1\over 6}
\sum_{{\sigma_{3},\sigma_{4}, \sigma_{5},\sigma_{6}}\atop
{z_{3},z_{4},z_{5},z_{6}}}
[&\rho_{6}(\sigma_{1},\sigma_{2},\sigma_{3},\sigma_{4},\sigma_{5},\sigma_{6},
z_{1},z_{2},z_{3},z_{4},z_{5},z_{6})\cr +\
&\rho_{6}(\sigma_{3},\sigma_{2},\sigma_{1},\sigma_{4},\sigma_{5},\sigma_{6},
z_{3},z_{2},z_{1},z_{4},z_{5},z_{6})\cr +\
&\rho_{6}(\sigma_{3},\sigma_{4},\sigma_{1},\sigma_{2},\sigma_{5},\sigma_{6},
z_{3},z_{4},z_{1},z_{2},z_{5},z_{6})\cr +\
&\rho_{6}(\sigma_{3},\sigma_{4},\sigma_{5},\sigma_{2},\sigma_{1},\sigma_{6},
z_{3},z_{4},z_{5},z_{2},z_{1},z_{6})\cr +\
&\rho_{6}(\sigma_{3},\sigma_{4},\sigma_{5},\sigma_{6},\sigma_{1},\sigma_{2},
z_{3},z_{4},z_{5},z_{6},z_{1},z_{2})\cr +\
&\rho_{6}(\sigma_{1},\sigma_{6},\sigma_{3},\sigma_{4},\sigma_{5},\sigma_{2},
z_{1},z_{6},z_{3},z_{4},z_{5},z_{2})]\cr }} and
\eqn\ttrace{\eqalign{\rho_{1A}(\sigma_{1},z_{1})&\equiv
\sum_{\sigma_{2},z_{2}} \rho_{2}(\sigma_{1},\sigma_{2},z_{1},z_{2}) \,
, \cr \rho_{1B}(\sigma_{2},z_{2})& \equiv \sum_{\sigma_{1},z_{1}}
\rho_{2}(\sigma_{1},\sigma_{2},z_{1},z_{2})\, .\cr} } 
The two-point function $\rho_2$ is constructed to be invariant under
rotations of the hexagon by $2\pi/3$, i.e. by those rotations which
leave the two subsets A and B invariant. Using the CVM equation below,
this will in turn imply that the six-point function $\rho_6$ itself
has this $2\pi/3$ rotational symmetry.  In \ttrace\ we have introduced
two site density matrices, $\rho_{1A}$ and $\rho_{1B}$, corresponding
to the two subsets A and B.

Noting that the numbers $N_H$, $N_P$, $N_{TA}$ and $N_{TB}$ of
hexagons, pairs and triangles of the subsets A and B, respectively,
satisfy $N_P/N_H=3$, $N_{TA}/N_H=N_{TB}/N_H=1$, one finds the
following approximate CVM free energy {\it per hexagon}: 
\eqn\fenergy{
\eqalign{&f(\rho_{6}(\{\sigma_{i}\},\{z_{i}\}))=- K
\sum_{\{\sigma,z\}}\sigma_{1}\sigma_{2} (1+2z_1z_2)
\rho_{2}(\sigma_{1},\sigma_{2},z_{1},z_{2}) +\lambda
\sum_{\{\sigma,z\}} \rho_6\cr &+\sum_{\{\sigma,z\}}(\rho_{6}\ln
\rho_{6})-3\sum_{\{\sigma,z\}}(\rho_{2}\ln \rho_{2})
+\sum_{\{\sigma,z\}}(\rho_{1A}\ln \rho_{1A})
+\sum_{\{\sigma,z\}}(\rho_{1B}\ln \rho_{1B}) , \cr 
}} 
which is a functional of $\rho_6(\{\sigma_i\},\{z_i\})$ only 
(by implicit use of
Eqs.\trace\ and \ttrace ), to be minimised with respect to
$\rho_6(\{\sigma_i\},\{z_i\})$.  Here $\lambda$ is a Lagrange
multiplier which ensures the normalisation of
$\rho_{6}(\{\sigma_i\},\{z_i\})$ to one.  The derivative with respect
to a generic element of $\rho_{6}$ yields the stationarity conditions
\eqn\stationary{
\eqalign{\rho_{6}(\{\sigma_{i}\},\{z_{i}\})&=\exp[-\lambda +
{K \over 2}\sum_{i=1,6}
(1+2z_{i}z_{i+1})\sigma_{i}\sigma_{i+1}]  \cr
&\times
[\rho_{2}(\sigma_{1},\sigma_{6},z_{1},z_{6})\rho_{2}(\sigma_{1},
\sigma_{2},z_{1},z_{2})
\rho_{2}(\sigma_{3},\sigma_{2},z_{3},z_{2}) \cr
&\ \ \rho_{2}(\sigma_{3},\sigma_{4},z_{3},z_{4})
\rho_{2}(\sigma_{5},\sigma_{4},z_{5},z_{4})
\rho_{2}(\sigma_{5},\sigma_{6},z_{5},z_{6})]^{1/2}             \cr
&\times [\rho_{1A}(\sigma_{1},z_1)\rho_{1B}(\sigma_{2},z_2)
\rho_{1A}(\sigma_{3},z_3)\rho_{1B}(\sigma_{4},z_4)
\rho_{1A}(\sigma_{5},z_5)\rho_{1B}(\sigma_{6},z_6)]^{-1/3}     ,\cr
}}
with the convention $\sigma_7=\sigma_1$, $z_7=z_1$.

One can solve this set of equations, with the definitions of
Eqs. \trace\ and \ttrace , by iteration. Starting from some assumed form
for $\rho_{2}$ and iterating the above equation, $\rho_{6}$ converges
to a solution of \stationary\ which is, moreover, a local {\it
minimum} of the approximate free energy \fenergy .  At each step, the
normalisation condition is imposed by adjusting the Lagrange
multiplier $\lambda$.

In practice, we expect that the expectation values $P$, $O$ and $T$ of
Eqs. \OP\ are the same for negative $K$ when calculated on the subset
A only or the subset B only. This implies $\rho_{1A}(\sigma,z)
=\rho_{1B}(-\sigma,z)$. It is thus more convenient to use the variable
$\sigma_{\rm st}$, instead of $\sigma$, and to introduce a single one
point function $\rho_1^{\rm st}(\sigma_{\rm st},z)\equiv
\rho_{1A}(\sigma_{\rm st},z) =\rho_{1B}(-\sigma_{\rm st},z)$ which is
independent of the subset A or B.  Eq. \stationary , which is a set of
$96\times 4$ coupled equations, can be reduced to a much smaller set
of equations by introducing appropriate ratios of six-point functions
and noting that some of these ratios are identical. For instance:
\eqn\insta{\eqalign {{\rho_6(++-+--,++++++)\over
\rho_6(--+-++,++++++)} &={\rho_6(---+++,++++++)\over
\rho_6(+++---,++++++)}\cr &={\rho_2(-+,++)\over
\rho_2(+-,++)}\left({\rho_{1A}(+,+)\rho_{1B}(-,+) \over
\rho_{1A}(-,+)\rho_{1B}(+,+)}\right)^{1\over 3} . \cr}}

We can then use an appropriate parametrisation of the six-point
function in terms of a small number of independent ratios. For these
we get a much smaller closed set of coupled equations which can be
solved numerically by recursion. In practice, we used a set of 9
independent ratios.

\subsec{Results of the CVM for the 2-d (arbitrary $K$) and 3-d (positive
$K$) folding problems} Before we present our results for the
3-d folding problem at negative $K$, let us briefly recall the results
which have been obtained in \CGP\ , where the CVM was applied to the
2-d folding problem (i.e. to the planar 11-vertex submodel) with
arbitrary $K$, and those of \CGPII\ for the full 3-d folding problem in
the FCC lattice with positive $K$.

\fig{The phase diagram in the bending rigidity $K$ of the 2-d folding
problem.}  {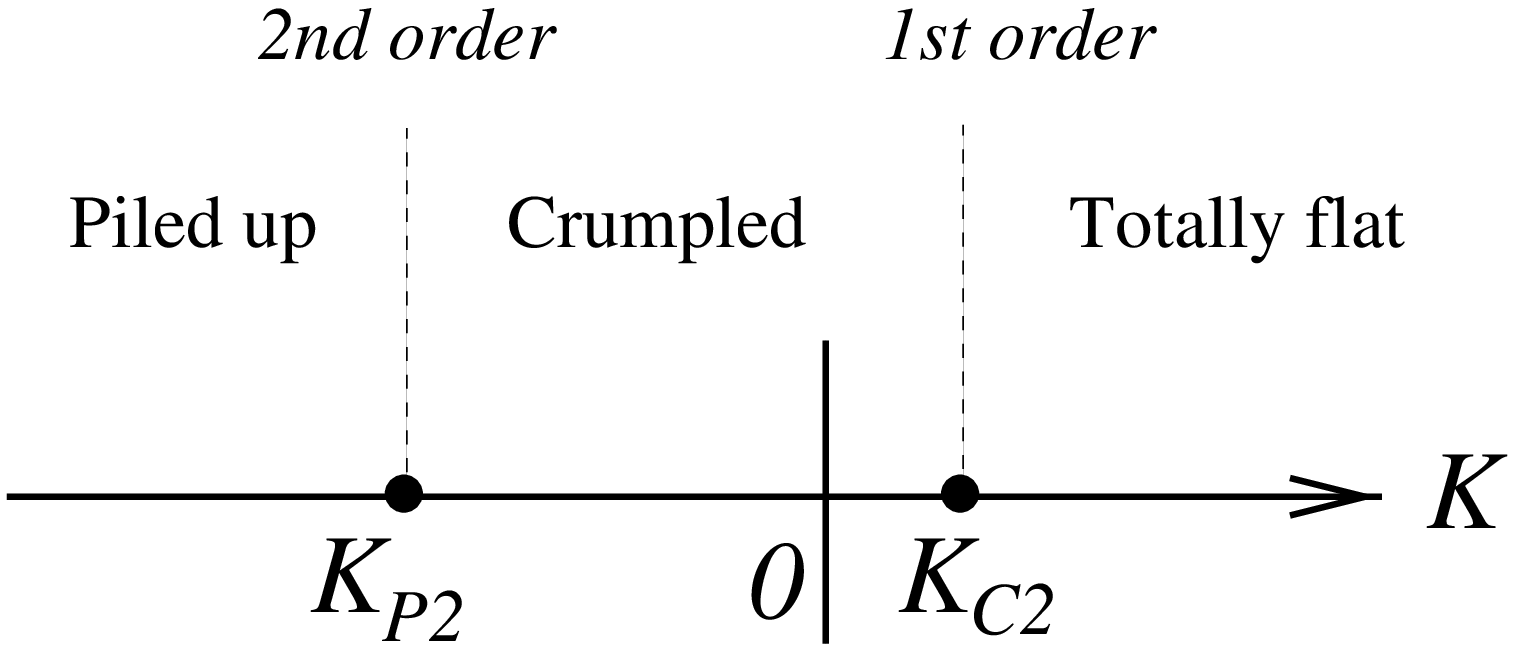}{4.0truein} \figlabel\phase 
For the 2-d folding problem, characterised in our notation by $z=+1$
everywhere, the CVM predicts two transitions, as depicted in
Fig. \phase .  A first order crumpling transition at $K_{C2}\sim
0.1013$ separates a large positive $K$ {\it totally flat} phase with
$\sigma=+1$ everywhere from a low $\vert K\vert $ crumpled phase with
$\langle \sigma \rangle=0$.  This latter phase also has $\langle
\sigma_{\rm st}\rangle=0$ and persists with decreasing $K$ to negative
values until $K_{P2}\sim -0.284$, where a second order piling
transition takes place.\foot{An exact formula for $K_{P2}$ can be
found in \DGM .}  For $K<K_{P2}$, the system develops a non-zero
value of $\langle \sigma_{\rm st}\rangle$, indicating a tendency for
triangles to pile on top of each other.  At $K=0$, the number
of folded states for $N$ triangles is found to grow like $q^N$ with
$q=\sqrt{13}/3\sim 1.202$, to be compared with the exact value $q=
\sqrt{3} \Gamma(1/3)^{3/2}/2\pi \sim 1.208$.  Note that this phase
diagram is also found by another approach to the problem using a
transfer matrix formalism \DIGG .

We stress here that, thanks to the duality transformation, this phase
diagram applies to the 11-vertex model describing folding on a single
tetrahedron.  The only modification is that the energy difference
between the two types of allowed folds, which is $K-(-K)=2K$ in the
planar problem, is now $-K/3-(-K)=2K/3$. The phase diagram of the
tetrahedral folding problem is thus exactly the same as that of the
planar folding problem with $K\to 3K$.  For instance the piling
transition now occurs at $K\sim -3\times 0.284= -0.852$.

For the 3-d folding problem in the FCC lattice, the CVM method has been
used in \CGPII\ for positive $K$ only. There it was found that, as in
the 2-d case, a first order crumpling transition takes place, now at a
larger value $K_{C3}\sim 0.1855$ between a large $K$ totally flat
phase with $z=\sigma=+1$ everywhere, and a crumpled state with
$\langle z\rangle=\langle \sigma \rangle=0$.  A remarkable difference
is that the low $K$ crumpled state now has $\langle \sigma_{\rm
st}\rangle $ non-zero. For instance, at $K=0$, one gets $\langle
\sigma_{\rm st}\rangle \simeq 0.87456$.  As mentioned at the beginning
of this section, this indicates a strong preference for the lattice to
wrap on octahedra. This non-zero value of the staggered $\sigma$
variable is not surprising in the sense that the 3-d folding problem in
the FCC lattice may be formulated approximately as a 2-d folding
problem with a staggered magnetic field (see \BDGG ). Furthermore, a
comparison, at $K=0$, of the entropy factor $q$ for the full
3-d folding problem (estimated to be 1.428 by the CVM and 1.43 by a
transfer matrix approach \BDGG ) with that of the single octahedron
folding problem ($q=\sqrt{2}\sim 1.414$) shows that the wrapping of
octahedra is the entropically dominant folding process at zero bending
rigidity.

\subsec{ 3-d folding at negative $K$}

We now present our results for the CVM applied to the full 3-d folding
problem in the FCC lattice, with {\it negative} $K$. Our main result
is the existence of two separate geometrical transitions with
increasing $\vert K\vert$: a first discontinuous transition at
$K=K_{G3}\sim -0.294$ separates a phase where the triangular lattice
is preferentially wrapped around octahedra from a phase where it is
preferentially wrapped around tetrahedra.  A second continuous
transition at $K=K_{P3}\sim -0.852$ separates this latter phase from a
phase of complete folding of the lattice on top of a single triangle.

\fig{The average proportions of the four types of folds, as a function
of $K$.}{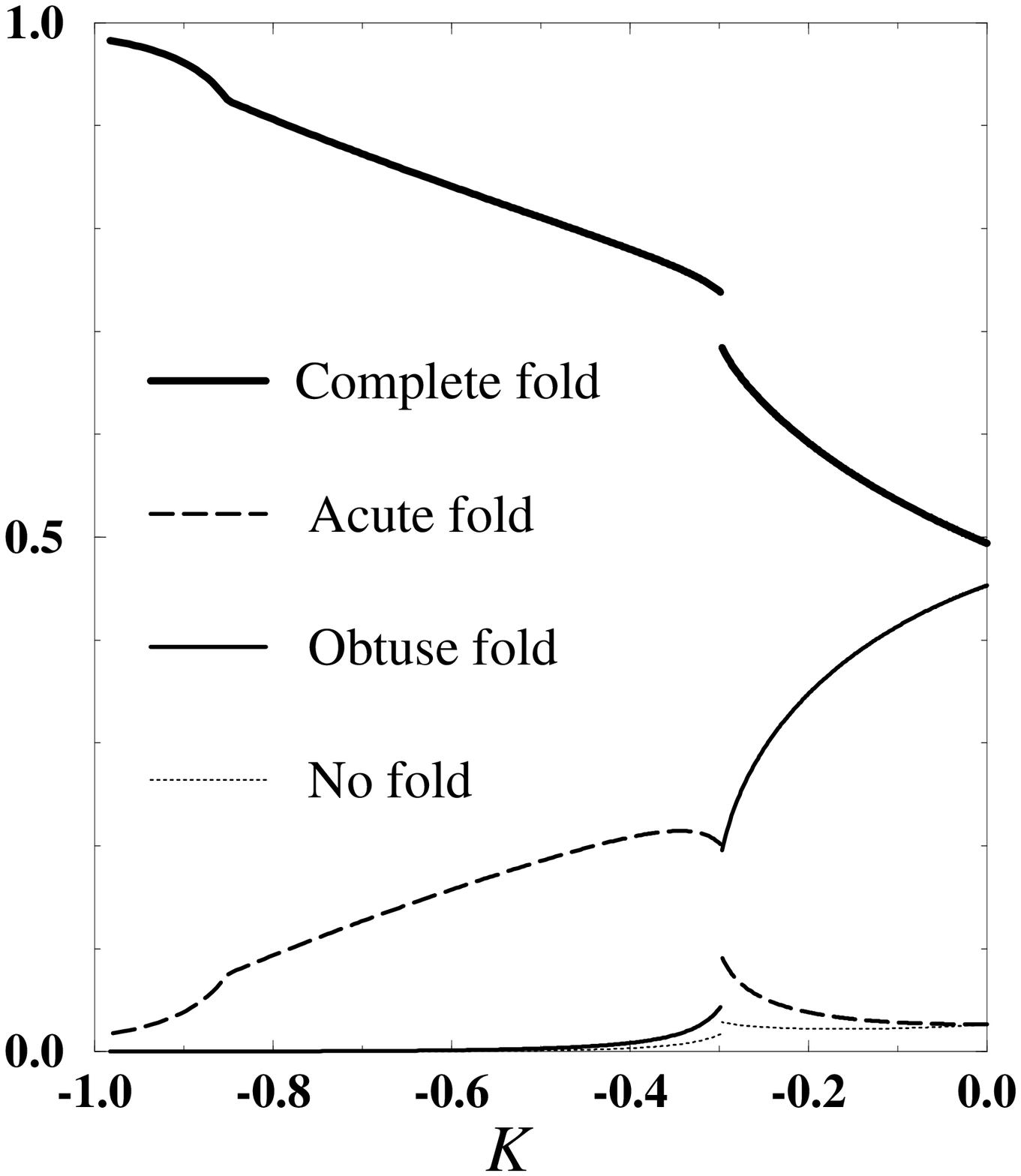}{4.0truein} \figlabel\prop The first sign of these
transitions can be seen directly by looking at the evolution with $K$
of the relative proportions of the four types of folds averaged over
the folded states of the lattice.  These proportions are displayed in
Fig. \prop .  At $K=0$, the fraction of acute folds and no-folds is
identical, as a consequence of duality.  Together they represent,
however, only $5\%$ of all folds and the statistics is dominated by
complete and obtuse folds, indicating a phase of wrapping of
octahedra.  This domination of complete and obtuse folds persists in
the range $K_{G3}<K\le 0$. At $K=K_{G3}$, all proportions show a
discontinuity.  For $K<K_{G3}$, the statistics is now dominated by
complete and acute folds, indicating a phase of wrapping of
tetrahedra.  The proportion of the other types of folds becomes
essentially zero for $K\sim -0.6$. The proportion $p(K)$ of complete
folds (respectively $1-p(K)$ of acute folds) then increases
(respectively decreases) smoothly with increasing $\vert K\vert$. This
domination of complete folds is enhanced at $K=K_{P3}$, where a
discontinuity in the slope of $p(K)$ signals a continuous transition.
\fig{The octahedral order parameter $O$, tetrahedral order parameter
$T$ and planar order parameter $P$ as a function of the bending
rigidity $K$.}{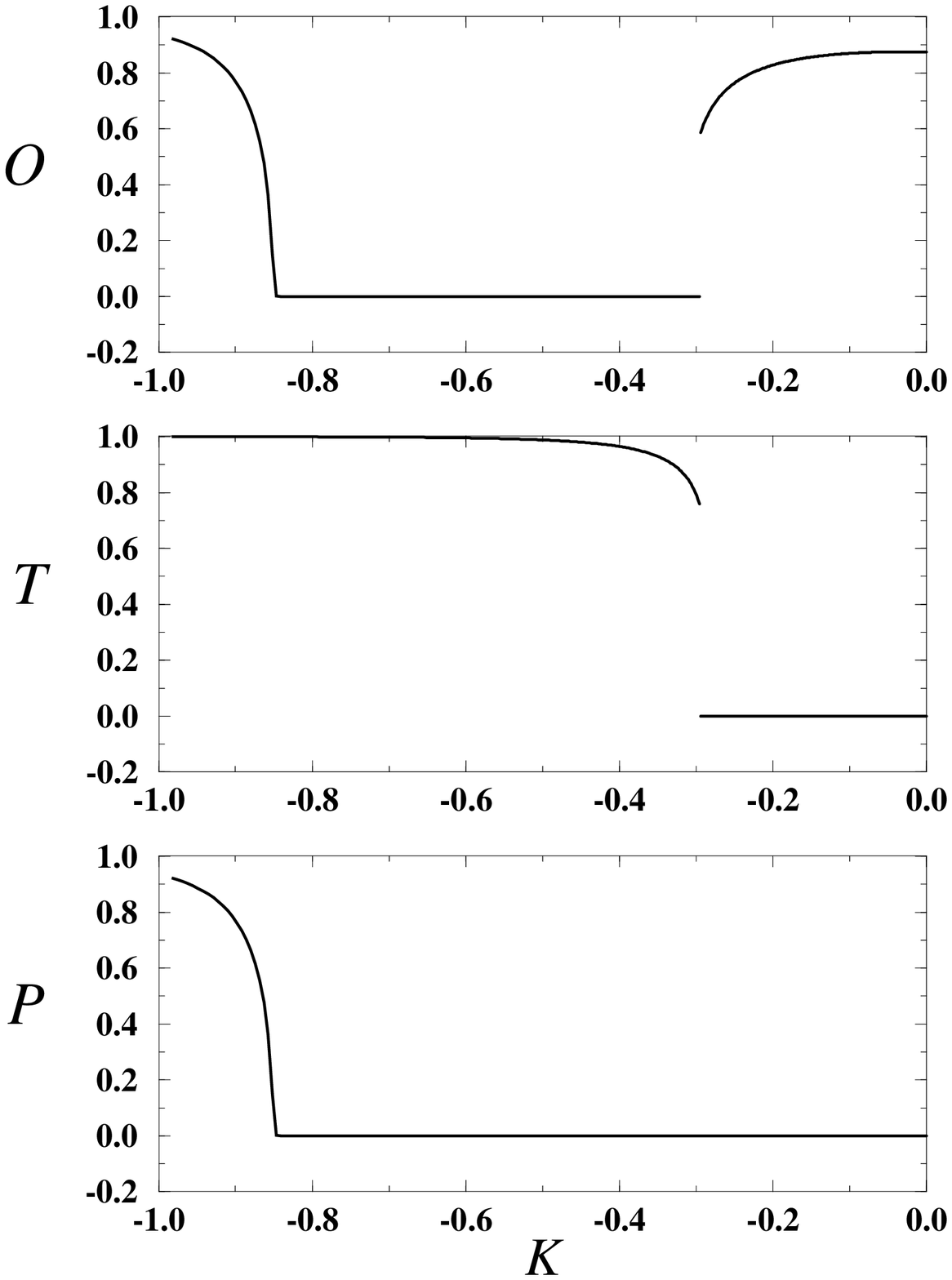}{4.0truein} \figlabel\OPfig A more detailed
description is provided by examining the three order parameters $O$,
$T$, and $P$ defined in \OP , whose values, displayed in Fig. \OPfig ,
clearly indicate three separate phases.  At low $\vert K\vert$ the
system is preferentially wrapped around octahedra, i.e. is in a phase
with a non-zero value of the octahedral order parameter $O$, while the
other two order parameters $T$ and $P$ remain zero.  For intermediate
$\vert K\vert$, the system is preferentially wrapped around
tetrahedra, i.e. is in a phase with a non-zero value of the
tetrahedral parameter $T$, while the two other order parameters
vanish.  The geometrical transition at $K=K_{G3}\sim -0.294$ between
these two phases is first order.  At large values of $\vert K\vert$,
the three order parameters are non-zero with $T\simeq +1$ and $O\simeq
P$. This clearly indicates a phase where the triangles are
preferentially piled up on top of each other. In practice, the
tetrahedral order parameter becomes indistinguishable from its
saturated value $T=+1$ for $\vert K\vert > 0.6$, so that at large
$\vert K\vert$, the 96-vertex model becomes indistinguishable from the
pure tetrahedral 11-vertex model (defined by $\sigma_{\rm st}=z$),
which satisfies $O=P$. It is therefore not surprising to recover the
second order piling transition of the 11-vertex model, which, as
explained above, occurs at $K_{P3}\sim -3\times 0.284=-0.852$.

\fig{The phase diagram in the bending rigidity $K$ of the 3-d folding
problem in the FCC lattice.}  {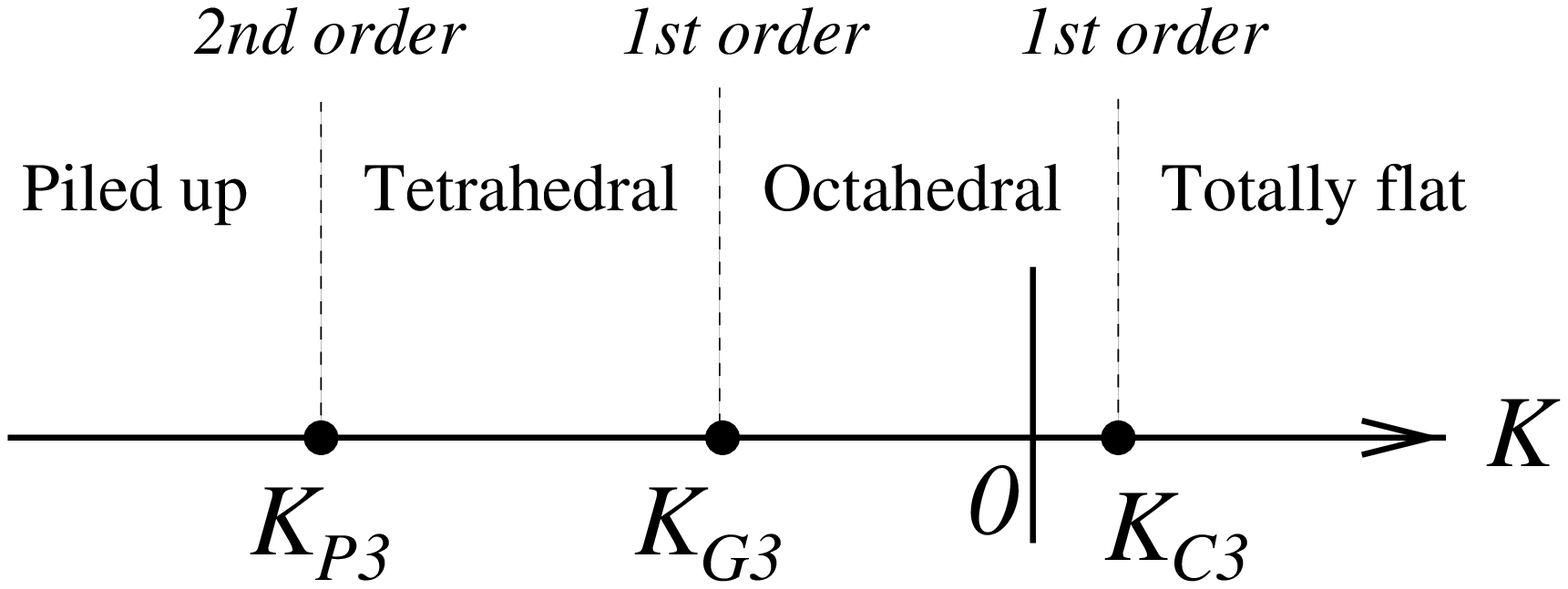}{4.0truein}
\figlabel\phaseII 
The final phase diagram of the 3-d folding problem of
the triangular lattice in the FCC lattice is shown in Fig. \phaseII . 
When compared with that of the 2-d planar folding problem, displayed
in Fig. \phase , we see that the crumpled phase has been replaced by
two geometrical octahedral and tetrahedral phases, separated by a
first order transition.  Note that the octahedral phase persists at
positive $K$ until the transition point $K_{C3}$ where the totally
flat phase is recovered. The passage from the totally flat phase to
the piled up phase, which involved two transitions in the 2-d case, now
requires three transitions.

It is interesting to compare our model with the Ashkin-Teller model
\Baxbook\ defined in terms of two {\it unconstrained} $Z_2$ variables
$\sigma_{\rm st}$ and $z$ with pair energy: \eqn\AT{E=-K_O
(\sigma_{\rm st})_1(\sigma_{\rm st})_2-K_P z_1 z_2 -K_T (z\sigma_{\rm
st})_1(z\sigma_{\rm st})_2 } involving three energy parameters
$K_{O,P,T}$. Note that, despite the apparent asymmetry, the three
$Z_2$ variables $\sigma_{\rm st}$, $z$ and $z\sigma_{\rm st}$ are
actually exactly on the same footing: for any two of them, the third
one is simply the product of these two. Our energy \ener\ corresponds
to a trivial case $(K_O,K_P,K_T)=(-K/3,0,-2K/3)$ where one of the
coupling constants vanishes. The {\it unconstrained} system then
translates into two independent Ising models with two continuous
transitions with increasing $\vert K\vert $ describing the successive
symmetry breakings of first $\langle \sigma_{\rm st}z\rangle$ at
$K=-3K_c/2$ and then of $\langle\sigma_{\rm st} \rangle$ at
$K=-3K_c$, while $\langle z\rangle=\langle \sigma_{\rm st} \rangle
\langle z\sigma_{\rm st}\rangle$ for all $K$. Here $K_c=-(1/2)\ln
(2-\sqrt{3}) \simeq 0.658$ is the critical Ising coupling on the
honeycomb lattice. In particular, at low $\vert K\vert$, all order
parameters vanish. We see that the constraints \firstrule\ and
\secondrule\ on the $\sigma_{\rm st}$ and $z$ variables modify this
picture by replacing the first continuous transition by a
discontinuous one and by giving rise to a low $\vert K\vert $ phase
where $\langle \sigma_{\rm st}\rangle$ is non-zero.  In this phase,
the origin of the symmetry breaking in the variable $\sigma_{\rm st}$
is clearly entropic. Indeed it allows for a gain in entropy from the
$z$ variable as the second basic folding rule is less constraining
when the $\sigma$ variables alternate on the lattice.

\fig{The two-point functions $O2$, $T2$ and $P2$ and the energy $e$,
free energy $f$ and entropy $s=e-f$ per hexagon, as a function of the
negative bending rigidity $K$.}  {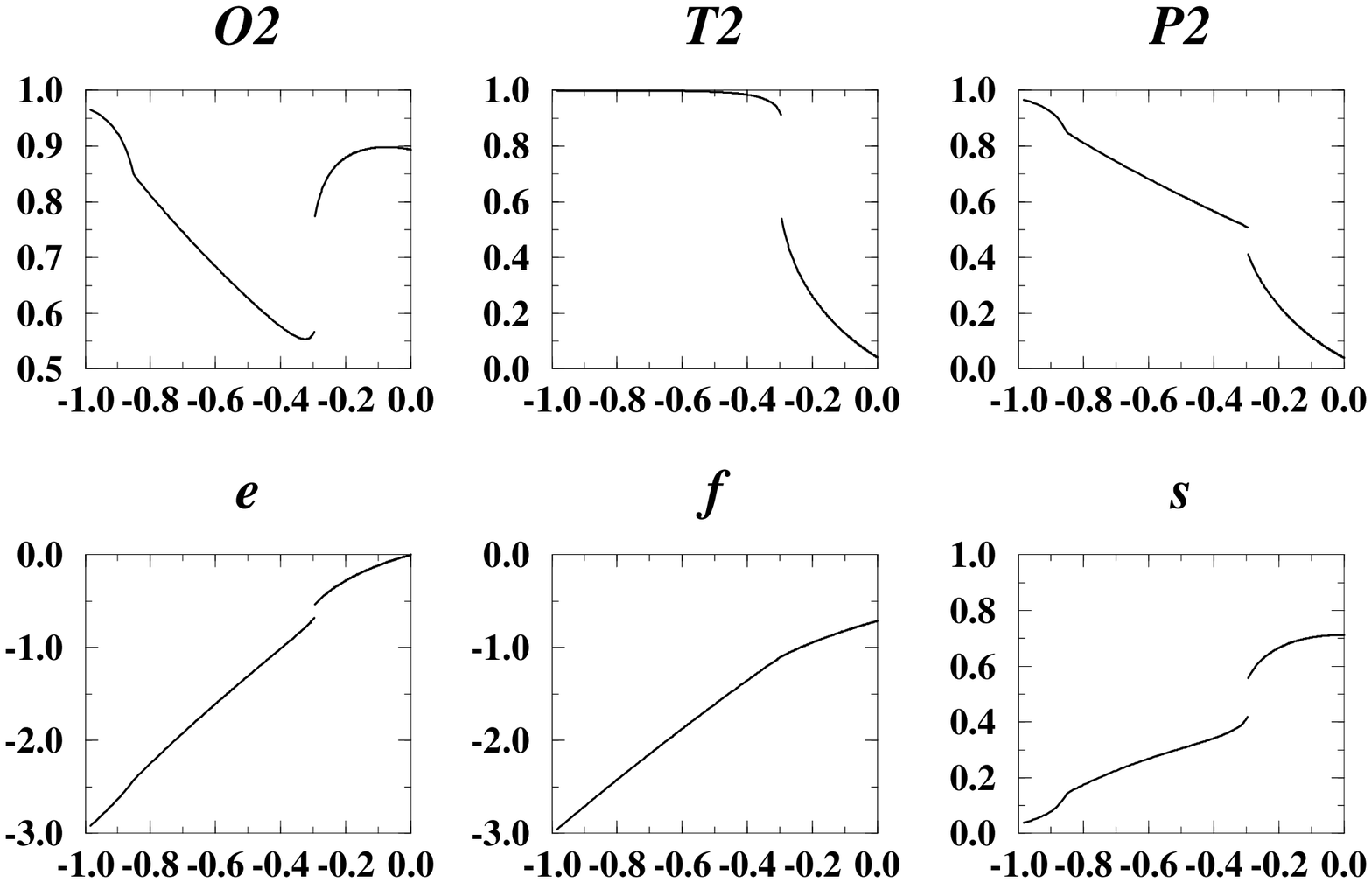}{5.0truein} \figlabel\cor
Since we can extract the two-point function from the CVM, we have
access to the nearest-neighbours correlations $O2$, $T2$ and $P2$
defined in \TWO , as well as to the average energy {\it per hexagon}:
\eqn\enperh{e=K\,(O2+2T2).}  We also have access to the free energy
$f$ per hexagon and to the entropy part $s=e-f$. These quantities are
displayed in Fig. \cor\ as a function of $K$.

\subsec {The transfer matrix approach}

The CVM gives only an approximation to the exact statistics of the
problem and it is difficult to estimate quantitatively possible
systematic errors of the method. In particular, due to the truncation
of the free energy functional to finite clusters, we cannot even
ensure that the CVM free energy is bounded from below by the true free
energy. It is thus important to compare the CVM results with those of
other methods. In the case of 2-d folding, a formulation of the problem
was given in terms which allows for the use of a transfer matrix
method \DIGG , leading to a very satisfactory comparison with the CVM
approach. The same method was generalised for the 3-d folding problem
at $K=0$ in \BDGG .  In this section, we present the results of a
transfer matrix study of the 3-d folding problem at negative bending
rigidity $K$.

In the transfer matrix formalism, the folding partition function
$Z_L(K)$ of an infinite strip with a finite width $L$ and with
rigidity $K$ is given by \eqn\trpart{Z_L(K) = \lim_{n\to\infty} [\Tr (
T_L(K) ^n)] ^{1/n} ,} where $T_L(K)$ is the row to row transfer matrix
of the system. The state of a row is specified by the $2\times 2L$
triangle variables $\sigma_i$ and $z_i$ in the row. We use free
boundary conditions which allow folds to escape from the strip at its
boundary. In the folding problem, a transfer matrix element from the
lower row $\{\sigma,z\}$ to the upper row $\{\sigma',z'\}$ \eqn\trelt{
T_{L \{\sigma,z\},\{\sigma',z'\} }= {\epsfxsize=7.truecm
\epsfbox{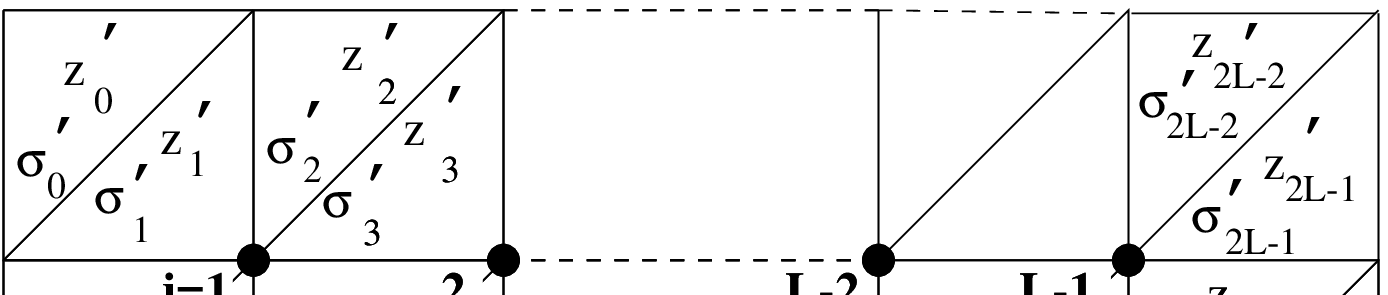}} }
\bigskip \bigskip \noindent is non-zero {\it iff} the two basic folding
rules \firstrule\ and \secondrule\ are satisfied around each of the
vertices $j=1,\dots,L-1$; if so, it is equal to $\exp (-E)$, where $E$
is the folding energy \ener, proportional to $K$, for all pairs of
neighbouring triangles in the two rows.  This method has been
extensively described in \BDGG\ for the 3-d folding {\sl without}
rigidity ($K=0$).  The question of the rigidity has been addressed in
\DIGG\ for the 2-d folding problem. The reader is referred to these
articles for further details on the method.

As usual, we must extract the largest eigenvalue (and its eigenvector)
of the matrix $T_L(K)$. The main difficulty is that the size of this
matrix increases exponentially with $L$ like $4^{2L} \times
4^{2L}$. Also, due to the chirality in the representation of the
triangular lattice as a square lattice in \trelt, the transfer matrix
is not directly symmetric, which requires the use of both left and
right eigenvectors.  Fortunately, the vanishing of all elements which
do not satisfy the basic folding rules makes the matrix very sparse.
With the Cray C94 of CEA-Grenoble, we have been able to reach $L=6$
with one Gigabyte of memory.  By comparison, in 2-d folding, the size
of the matrix was $2^{2L} \times 2^{2L}$ and the width $L$ could be
doubled. These computations are exact, in the sense that the precision
is limited by the numerical precision of the computer; no
approximations are made with this method. Of course, the infinite
lattice folding is recovered only in the limit where $L\to\infty$.

Strictly speaking, an infinite strip with finite width is a
one-dimensional system which has no phase transition and all order
parameters vanish due to the global symmetry under reversal of all
spins.  The order parameters, however, can be estimated by weakly
breaking the symmetry by first fixing the $(\sigma,z)$ variables of,
say, the most central triangle of a row $i=L$ and by averaging each
order parameter on the other triangles.  In practice, it is useful to
restrict oneself to triangles $j$ with $|j-i|>1$, i.e.~to triangles
which are not nearest neighbours of $i$. In fact, the contribution of
neighbouring triangles is less significant because the short range
correlations are always large even in a disordered phase.  To summarise, 
we measure $X_L = 1/(2L-3) \sum_{j:|j-L|>1} \langle x_L x_j\rangle$
as an estimate for the order parameter $X=\langle x \rangle$ to be 
considered ($x=\sigma_{\rm st},z\sigma_{\rm st}$ or $z$, i.e. 
$X=O,T$ or $P$).

Using the Pad\'e-Shanks transformation (see Ref.~\BREZ ), the finite
size effects can be reduced.  Given three values $X_4, X_5, X_6$
obtained for sizes $L=4,5,6$, and supposing the convergence is exactly
exponential, the extrapolated value is $X_{\infty} = (X_4 X_6 - X_5^2)
/ (X_4 + X_6 - 2 X_5)$.

\fig{The results of the transfer matrix method for the tetrahedral
order parameter $T$ as a function of (negative) $K$. The figure
displays the results for strips of size $L=3,4,5,6$ as well as the
extrapolation to infinite $L$ (see text). By comparison, the
corresponding result of the CVM is also indicated.}
{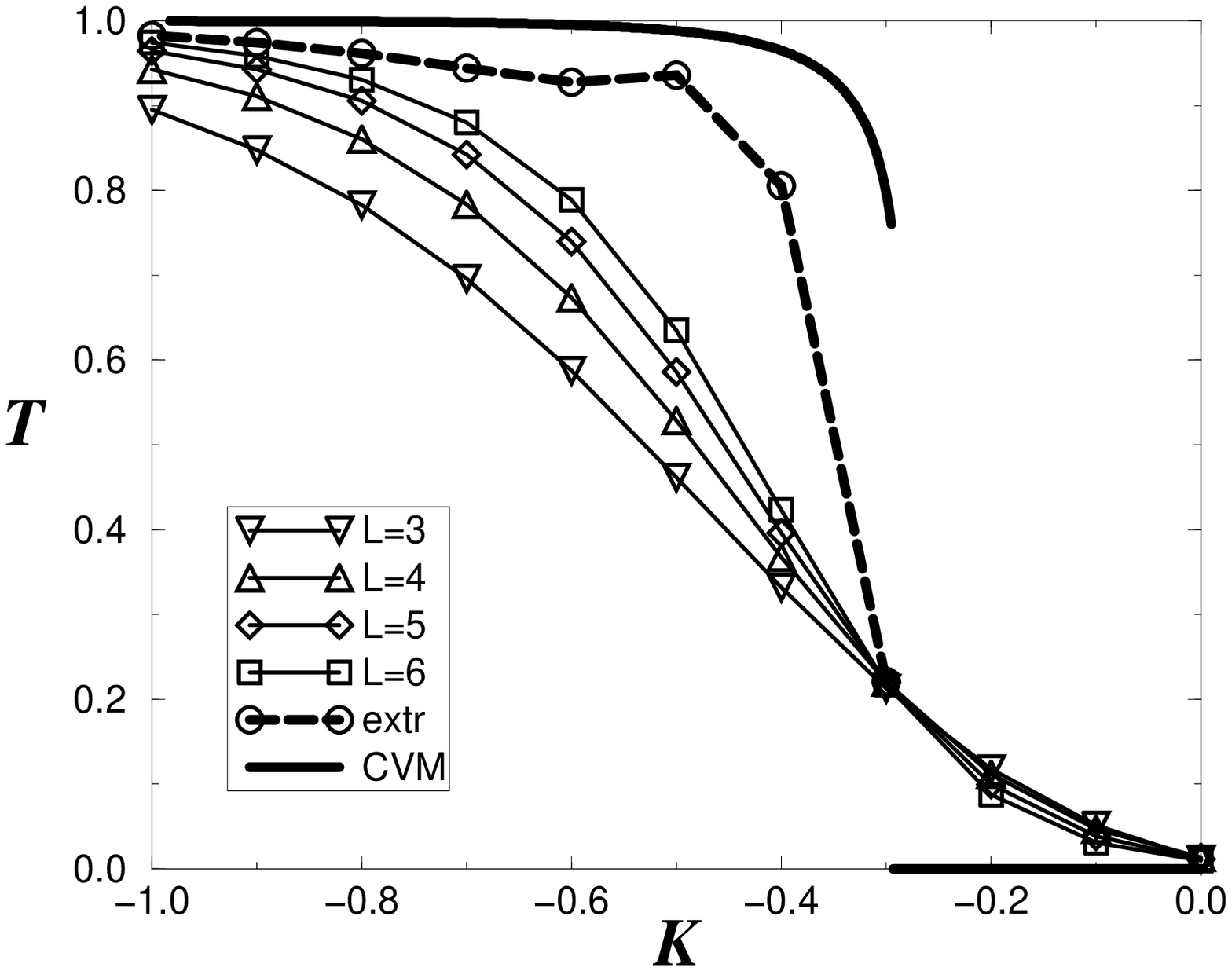}{4.0truein} \figlabel\trans In Fig.~\trans , the tetrahedral
order parameter $T$, estimated as described above, is plotted as a
function of the negative rigidity $K$.  For finite width $L=6$, the
first order transition predicted by the CVM is smoothed by finite size
effects. The extrapolation $L\to\infty$, however, clearly shows a
rapid change around $K=-0.3$.  Unfortunately, the maximum size $L=6$
is too small to allow a precise determination of $K_{G3}$ or even the
order of the transition. We can still estimate $K_{G3}$ roughly by
considering the point where all curves at finite $L$ seem to
intersect. This leads to $K_{G3}\sim -0.3$, in good agreement with the
CVM estimate $K_{G3}\sim -0.294$.

For the other order parameters ($P$ and $O$), the curves are smoother
and the predicted transitions less visible. Still, the variations of
these order parameters with $K$ are in agreement with the infinite
system limit expected from the CVM results. For instance, we see an
intermediate $\vert K\vert$ regime with a low value of $O$ surrounded
by a small and large $\vert K\vert $ regime with a significantly
larger value of $O$, as expected in the presence of an intermediate
tetrahedral phase.
 
Clearly, the exact transfer matrix method is more difficult in the
3-d folding case than in the 2-d folding case, due to the higher number
of states.  In the 2-d case, the first signal of the continuous piling
transition at $K_{P2}$ did not occur until system width $L=8$. Such a
large system size is impractical in the 3-d problem.  In this context
we can conclude that the transfer matrix results do not contradict in
any way those of the CVM, which in turn appears to be a much more
powerful approach.

\newsec {External fields}

In this section, we briefly discuss the consequence of adding external
fields coupled to our $Z_2$ spin variables. The results of this
section are again obtained in the framework of the CVM. We add to the
pair energy \ener\ the following external field contribution:
\eqn\extfield{E_H=-H_O \sigma_{\rm st} -H_T z\sigma_{\rm st} -H_P z \
,} with three external fields $H_O$, $H_T$ and $H_P$ coupled to the
order parameters $O=\langle \sigma_{\rm st}\rangle$, $T=\langle
z\sigma_{\rm st} \rangle$ and $P=\langle z\rangle$, respectively.

For simplicity, we will discuss in the following the consequence on
the negative $K$ phase diagram of having one of these fields non-zero,
while the two others are kept equal to zero.

As explained in Section 4.3, in the absence of external field, the
system was found to have three phases for negative $K$. The nature of
each phase can be characterised by specifying which of the three order
parameters $O$, $T$ and $P$ are zero or not.
\item{(i)} The low $\vert K\vert$ octahedral phase has $O\ne 0$, $P=T=0$.
\item{(ii)} The intermediate $\vert K\vert$ tetrahedral phase has
$O=0$, $P=0$ and $T\ne 0$.  
\item{(iii)} The large $\vert K\vert$ piled up phase has $O\ne 0$, $P\ne 0$ 
and $T\ne 0$.

We find that the nature of a phase is {\it unchanged} by the addition
of a (small enough) external field $H_X$ if the order parameter
$X$ was non-zero in this phase. Here $X$ stands for $O$, $T$ or $P$.
On the other hand, if $X$ happened to be zero in a phase, then
the action of $H_X$ is to make all order parameters non-zero.

For instance, the piled up phase, which has its three order parameters
non-zero, persists in the presence of external fields.  In the
presence of a non-zero value of $H_O$, the octahedral phase is
preserved but the continuous piling transition is {\it
suppressed}. The original tetrahedral and piled up phase merge into a
single phase with non-zero order parameters $O$, $T$ and $P$. This
phase can be viewed as a tetrahedral phase with a smoothly increasing
piled up order with increasing $\vert K\vert$. The first order
transition, now directly from the octahedral phase to this phase, is
slightly shifted to larger values of $\vert K\vert$.  In the presence
of a non-zero value of $H_T$, the tetrahedral and piled up phases are
not modified and the second order transition is preserved at its
original value of $K$. This is not surprising since $T$ was already
saturated to $T\simeq 1$ in this regime.  On the other hand, the
system develops non-zero order parameters $T$ and $P$ in the low
$\vert K\vert$ region, indicating the appearance of a piled up order
in the original octahedral phase.  The first order transition from
this phase to the tetrahedral phase is {\it preserved} (for small
enough $H_T$ at least), although slightly shifted to smaller values of
$\vert K\vert$.  Finally, in the presence of a non-zero value of
$H_P$, the three order parameters become non-zero over the whole
negative $K$ range. The second order transition is {\it suppressed}
while the first order transition now takes place between two phases of
the same nature, discriminated only by which fraction of acute or
obtuse folds is larger.

\fig{The variation with the negative bending rigidity $K$ of the three
order parameters $O$, $T$ and $P$ in the presence of a small applied
external field. The first (respectively second and third) column
corresponds to the application of a positive external field $H_O$
(respectively $H_T$ and $H_P$). The dashed vertical line indicates the
location of the first order geometrical transition in the absence of
external fields.  }{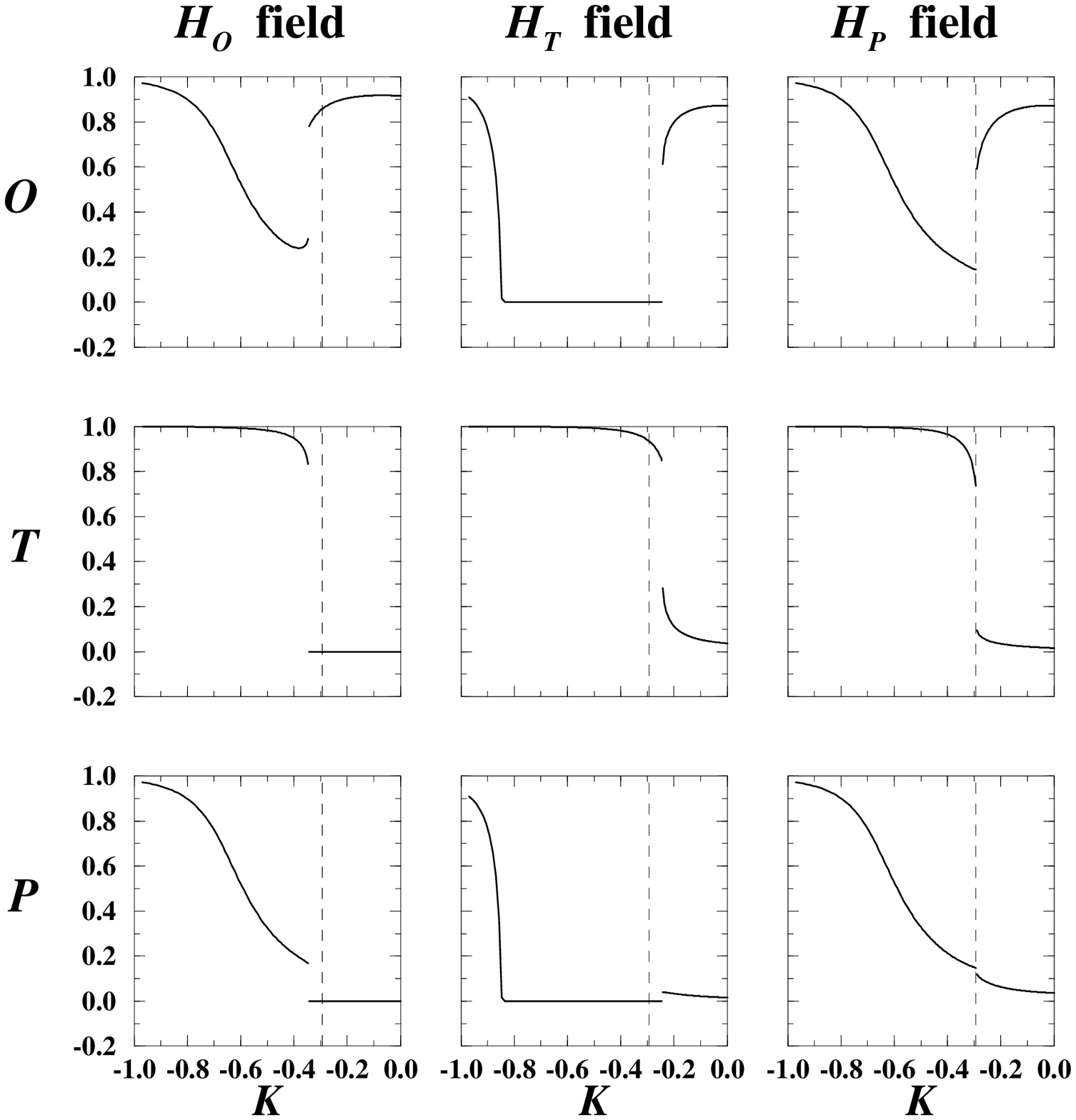}{5.0truein} \figlabel\withfields

The above discussion is illustrated in Fig. \withfields\ where the three
order parameters $O$, $P$ and $T$ and represented as a function of $K$
in the presence of an applied external field $H_O$, $H_T$ or $H_P$.  

\newsec {Discussion}

In this paper, we have studied the folding of the 2-dimensional
regular triangular lattice in the 3-dimensional regular
Face-Centred Cubic lattice, a problem which can be expressed as a
96-vertex model on the triangular lattice.  We considered the case of
negative bending rigidity $K$, where folds are favoured. A special
role is played by three classes of elementary objects in the FCC,
namely elementary octahedra, tetrahedra and triangles. Indeed, instead
of randomly wandering over all sites of the FCC lattice, the system
prefers to stay trapped on one class of these objects:
\item{-} At low $\vert K\vert$, the system is preferentially wrapped 
on octahedra, characterised by obtuse $109^\circ$ degree folds.
\item{-} At intermediate $\vert K\vert$, the system is preferentially 
wrapped on tetrahedra, with acute $71^\circ$ folds.
\item{-} At large $\vert K\vert$, the system is preferentially piled onto
single triangles, making complete $0^\circ$ folds.

\noindent This leads to three distinct phases, characterised by the
symmetry breakings of appropriate order parameters. These phases are
separated by two transitions, a discontinuous one between the
octahedral and the tetrahedral phase and a continuous one between the
tetrahedral and the piled up phase.  This study completes the previous
one at positive $K$ which found a first order crumpling transition
with a large positive $K$ completely flat phase.

Our results were obtained in the framework of the CVM approximation
and further corroborated by a transfer matrix approach.  At this
stage, one may wonder whether the particular form \energ\ of the
bending energy is crucial for our results. We expect that, as long as
the energy is a monotonic function of the angle of the fold which,
moreover, favours folding, the above scenario will continue to hold.
One may also question the validity of the CVM approximation in a very
constrained problem where folds are, moreover, expected to propagate
throughout the entire lattice, far beyond the size of the basic
cluster of the CVM. Actually, as explained in \DIG\ for the 2-d folding
problem, the necessity for folds to propagate through the lattice is
entirely encoded in the particular form of the 96 local hexagonal
configurations {\it and} the matching condition for the folds on links
which are common to two overlapping hexagons. These two ingredients
are properly taken into account by the CVM. We thus believe that this
method does indeed account for the propagation of folds and therefore
gives reliable results concerning the phase diagram and the
transitions separating distinct phases.  Nevertheless, as for any mean
field approach, the method gives only trivial critical exponents for
the continuous piling transition, whose true universality class is
therefore still to be determined.  A CVM approach with larger clusters
would not help to resolve this issue and we believe that the
restriction to hexagonal clusters is probably the most efficient
choice in terms of simplicity. On the other hand, a transfer matrix
method, which is exact for strips of finite size $L$ does not allow us
to reach large enough system sizes to allow a reliable continuum
($L\to \infty$) extrapolation.  It is thus a challenge to find
alternative methods for a more refined study of this problem.

Finally, the particular sequence of folding involving octahedra and
tetrahedra is clearly intimately linked to the structure of the FCC
target space where the 3-d rotational symmetry has been explicitly
broken to a smaller (cubic) symmetry. On the other hand, the piled up
phase with alternating normal vectors on neighbouring triangles is
likely to exist even if the target space is continuous.  One may
wonder whether the folding of the triangular lattice in a continuous
3-d space also gives rise to intermediate phases with particular broken
symmetries, not necessarily related of course to the FCC cubic
symmetry.
 
\bigskip
\noindent{\bf Acknowledgements}

The research of M.B. was supported by the Department of Energy, USA,
under contract N$^{\rm o}$ DE-FG02-85ER40237.  This work is partially
supported by the JSPS fellowship for Junior Scientists.  We thank
J.-M. Normand for a critical reading of the manuscript.
 
\listrefs
 
\end